\documentclass[challenges,article,submit,pdftex,moreauthors]{Definitions/mdpi} 

\usepackage[utf8]{inputenc}
\usepackage{newtxtext,newtxmath}
\usepackage[american]{babel}
\usepackage{csquotes}
\usepackage{multirow}
\usepackage{xurl}
\usepackage{booktabs}
\usepackage{tabularx,ragged2e}
\newcolumntype{L}{>{\raggedright\arraybackslash}X}
\newcolumntype{R}{>{\raggedleft\arraybackslash}X}
\usepackage{threeparttable}
\usepackage{stmaryrd}
\usepackage{float}
\usepackage{datetime}

\usepackage{titlesec}

\titleclass{\paragraph}{straight} 
\titleclass{\subparagraph}{straight} 
\titleformat{\paragraph}[runin]{\normalfont\normalsize\bfseries}{\theparagraph}{1em}{} 
\titleformat{\subparagraph}[runin]{\normalfont\normalsize\bfseries}{\thesubparagraph}{1em}{} 
\titlespacing*{\paragraph}{0pt}{3.25ex plus 1ex minus .2ex}{1em} 
\titlespacing*{\subparagraph}{\parindent}{3.25ex plus 1ex minus .2ex}{1em} 

\usepackage{pgfplots}
\pgfplotsset{compat=1.17}
\usepackage[outline]{contour}
\contourlength{0.2em}
\usepackage{tikz}
\usetikzlibrary{patterns}
\usepgfplotslibrary{fillbetween}
\usetikzlibrary{patterns}
\usetikzlibrary{positioning, calc}
\usetikzlibrary{arrows.meta}

\usepackage{siunitx}
\DeclareSIUnit\tx{tx}
\DeclareSIPrefix{\noop}{}{0} 
\usepackage[backend=biber, style=ieee]{biblatex}
\usepackage{acro}
\DeclareAcronym{PoW}{
  short = PoW,
  long  = proof-of-work
}
\DeclareAcronym{PoS}{
  short = PoS,
  long  = proof-of-stake
}
\DeclareAcronym{DLT}{
  short = DLT,
  long  = distributed ledger technology
}
\DeclareAcronym{TPS}{
  short = tps,
  long  = transactions per second
}
\DeclareAcronym{GHG}{
  short = GHG,
  long  = greenhouse gas
}
\DeclareAcronym{ASIC}{
  short = ASIC,
  long  = application-specific integrated circuit
}

\DeclareAcronym{PPA}{
  short = PPA,
  long  = power purchase agreement
}
\DeclareAcronym{L1}{
  short = L1,
  long  = layer-1
}
\DeclareAcronym{L2}{
  short = L2,
  long  = layer-2
}
\DeclareAcronym{IEA}{
  short = IEA,
  long  = International Energy Agency
}
\DeclareAcronym{GWP}{
  short = GWP,
  long  = global warming potential
}
\DeclareAcronym{BTM}{
  short = BTM,
  long  = behind-the-meter
}
\DeclareAcronym{FTM}{
  short = FTM,
  long  = front-of-the-meter
}
\DeclareAcronym{ROI}{
  short = ROI,
  long  = return on investment
}
\DeclareAcronym{CLR}{
  short = CLR,
  long  = controllable load resource
}
\DeclareAcronym{NCLR}{
  short = NCLR,
  long  = non-controllable load resource
}
\DeclareAcronym{RECs}{
  short = RECs,
  long  = Renewable energy certificates
}
\DeclareAcronym{GOs}{
  short = GOs,
  long  = guarantees of origin
}
\DeclareAcronym{ETFs}{
  short = ETFs,
  long  = exchange-traded funds
}
\DeclareAcronym{CO2}{
  short = CO2,
  long  = carbon dioxide
}
\DeclareAcronym{US}{
  short = US,
  long  = United States
}
\DeclareAcronym{EU}{
  short = EU,
  long  = European Union
}
\DeclareAcronym{VRE}{
  short = VRE,
  long  = variable renewable energy
}
\DeclareAcronym{BTC}{
  short = BTC,
  long  = Bitcoin Core
}
\DeclareAcronym{BMC}{
  short = BMC,
  long  = Bitcoin Mining Council
}
\DeclareAcronym{BECI}{
  short = BECI,
  long  = Bitcoin Energy Consumption Index
}
\DeclareAcronym{RES}{
  short = RES,
  long  = renewable energy sources
}
\DeclareAcronym{RE}{
  short = RE,
  long  = renewable energy
}
\DeclareAcronym{VOCs}{
  short = VOCs,
  long  = volatile organic compounds
}
\DeclareAcronym{IPCC}{
  short = IPCC,
  long  = Intergovernmental Panel on Climate Change
}
\DeclareAcronym{OSTP}{
  short = OSTP,
  long  = Office of Science and Technology Policy
}
\DeclareAcronym{LCOE}{
  short = LCOE,
  long  = Levelized Cost of Energy
}
\DeclareAcronym{DR}{
  short = DR,
  long  = demand-response
}
\DeclareAcronym{FLR}{
  short = FLR,
  long  = flexible load response
}
\DeclareAcronym{EACs}{
  short = EACs,
  long  = energy attribution certificates
}

\firstpage{1} 
\pubvolume{1}
\issuenum{1}
\articlenumber{0}
\pubyear{2023}
\copyrightyear{2023}
\datereceived{ } 
\daterevised{ } 
\dateaccepted{ } 
\datepublished{ } 
\hreflink{https://doi.org/} 

\Title{Can Bitcoin Stop Climate Change? Proof of Work, Energy Consumption and Carbon Footprint (SoK)}

\TitleCitation{Can Bitcoin Stop Climate Change? Proof of Work, Energy Consumption and Carbon Footprint (SoK)}

\Author{Juan Ignacio Ibañez $^{1,\dagger,\ddagger}$\orcidA{} and Alexander Freier $^{2,\ddagger}$}

\AuthorNames{Juan Ignacio Ibañez and Alexander Freier}

\AuthorCitation{Ibañez, J.; Freier, A.}

\address{%
$^{1}$ \quad UCL Centre for Blockchain Technologies, Catholic University of Córdoba, DLT Science Foundation\\
$^{2}$ \quad UCL Centre for Blockchain Technologies, Catholic University of Córdoba, Energiequelle GmbH}

\corres{Correspondence: j.ibanez@ucl.ac.uk}

\secondnote{These authors contributed equally to this work.}

\abstract{
Despite their potential in many respects, blockchain and \ac{DLT} technology have been the target of criticism for the energy intensity of the \ac{PoW} consensus algorithm in general and of Bitcoin mining in particular. However, mining is also believed to have the potential to drive net decarbonization and renewable penetration in the energy grid by providing ancillary and other services.
In this paper, we systematize the state of the art in this regard.
Although not completely absent from the literature, the extent to which \ac{FLR} through \ac{PoW} mining may support grid decarbonization remains insufficiently studied and hence contested.
We approach this research gap by systematizing both the strengths and the limitations of mining to provide \ac{FLR} services for energy grids.
We find that a net-decarbonizing effect led by renewable-based mining is indeed plausible.
}


\begin{document}




\section{Introduction}
\label{sec:introduction}

\subsection{Bitcoin, Proof of Work, and Energy Consumption}

In 2008, Satoshi Nakamoto \cite{Nakamoto2008Bitcoin:System} published the Bitcoin whitepaper, introducing a peer-to-peer payment system based on \ac{DLT} in the form of an append-only data structure called a blockchain. The core innovation of this system, known as \textit{Nakamoto consensus}, combined the blockchain with Dwork and Naor's \cite{Dwork1993PricingMail} \ac{PoW} to enable decentralized, censorship-resistant transactions.

Since its inception, Bitcoin\footnote{In its main fork, \ac{BTC}. The reader should note that multiple chains make claims to be the ``original" Bitcoin protocol. Where not indicated otherwise, "Bitcoin" in the remainder of this paper refers to \ac{BTC}.}
 has become the world's most prominent \ac{DLT}, with the market capitalization of its eponymous cryptocurrency bitcoin reaching a trillion US dollars \cite{Carter2021BitcoinZero,Williams2022Bitcoin:2022}. However, the energy-intensive nature of its \ac{PoW} mechanism has raised concerns about its \ac{GHG} emissions and ``social license to operate" \cite[3]{Rennie2021ClimateBitcoin} (see \ref{sec:bitcoin}).

The \ac{PoW} mechanism enables decentralized systems to achieve consensus while resisting Sybil attacks by tying digital identities (which would otherwise be nearly costless to create and could unduly influence the network) to a scarce resource, namely energy expenditure \cite{Piro2006DetectingNetworks,Trifa2014SybilAttack,Douceur2002TheAttack,Platt2021TheProof-of-Work,Ibanez2023TheExpansion}.\footnote{This is achieved by awarding the right to append a new block to the chain, along with a bitcoin subsidy known as the "block reward" and the option to request transaction fees to include transactions in one's block, to the node that finds a unique number called "nonce" through a process called "mining." Because the nonce can only be discovered by randomly guessing via the energy-intensive operation of hashing, the likelihood of successfully mining a bitcoin is proportional to the energy expended in mining, given similar hardware \cite{Carter2021BitcoinZero,Platt2021TheProof-of-Work,Ibanez2023TheExpansion}.} This design has proven robust, making Bitcoin the most successful and widely adopted cryptocurrency \cite{Carter2021BitcoinZero,Williams2022Bitcoin:2022}. The \ac{PoW} mechanism effectively supports censorship resistance and protocol stability, crucial for Bitcoin's purported role as a store of value \cite{Carter2021BitcoinZero}.

However, as bitcoin's price and interest in holding the asset have increased, so has energy consumption from mining. Advocates argue that a higher hashrate enhances protocol security \cite{Carter2021BitcoinZero}, while critics express concern over the significant carbon footprint\footnote{Carbon footprint is directly proportional to hashrate, all else being equal \cite{Oysti2021BitcoinConsumption}.} and its potential growth with further Bitcoin adoption.

\subsection{Flexible Load Response Through Cryptocurrency Mining}

Bitcoin proponents argue that despite its energy consumption and carbon footprint (which are also claimed to be overestimated, see \ref{sec:bitcoin}), the cryptocurrency could provide an environmental service through \ac{FLR} capabilities. This approach would support \ac{RE} profitability (and, thus, penetration), as well as methane onsite neutralization \cite{Roeck2022LifePlant,Dogan2022AreMethod}, and could theoretically result in \textit{net decarbonizing} additions of load.


Previous studies estimating Bitcoin's carbon footprint have been limited and contested (often not in a scientific fashion \cite{Corbet2021Bitcoin-energyEvidence}), while few have explored its potential synergy with \ac{FLR} or methane reduction (see \cite{Roeck2022LifePlant,Dogan2022AreMethod}). This paper aims to fill this gap by systematically reviewing the characteristics of Bitcoin mining and the renewable energy sector, potential complementarities, and limitations.

\subsection{Our contribution}

Our contribution consists of a comprehensive literature review, including academic papers, industry reports, and divulgation articles, supplemented by interviews with specialists and market players (see \ref{sec:acknowledgements}). We provide an overview of Bitcoin's environmental impact, discuss challenges in the renewable energy market, identify unique characteristics of Bitcoin mining relevant to decarbonization, and explore potential applications within the renewable energy sector. Additionally, we evaluate the positive effects of green Bitcoin mining, consider its limitations and challenges, and compare Bitcoin mining to alternative ancillary service providers.

The paper is structured as follows:
\begin{enumerate}
    \item Overview of literature on Bitcoin's environmental impact
    \item Discussion of the \ac{RE} market and decarbonization challenges
    \item Presentation of Bitcoin mining characteristics relevant to the decarbonziation goal
    \item Exploration of actual and potential Bitcoin mining applications within the \ac{RE} sector
    \item Analysis of positive effects of green Bitcoin mining
    \item Assessment of limitations and challenges of this approach
    \item Comparison of Bitcoin mining to other ancillary service providers
    \item Conclusion with a general discussion of findings, limitations, and future work
\end{enumerate}

\section{Bitcoin's Environmental Impact}
\label{sec:bitcoin}

\subsection{Public Discussion: Context}
\label{sec:context}

The primary environmental concern surrounding Bitcoin is its high energy consumption, even raised to Satoshi Nakamoto in 2010.\footnote{https://bitcointalk.org/index.php?topic=721.msg8114\#msg8114} This issue has contributed to slower cryptocurrency market penetration and even increased price volatility \cite{Buckland2021BitcoinPayments}. Consequently, energy consumption has become a central focus in academic research \cite{Dogan2022AreMethod}.\footnote{Despite numerous publications, this area remains under-researched \cite{Corbet2021Bitcoin-energyEvidence,Yazc2023AHardware}. Coinshares \cite{Coinshares2022TheImpact} claims that only Carter and Stevens' report \cite{Carter2021BitcoinZero} and its own use a sufficiently granular and accurate methodology.}

Bitcoin mining occurs globally, with miners seeking cost-effective locations offering the cheapest energy sources \cite{Jones2022EconomicGold,Kohler2019LifeMining,Oysti2021BitcoinConsumption,McCook2022DriversEmissions,Nikzad2022Techno-economicMining,Velicky2023RenewableBitcoin}.\footnote{See the discussion of \ac{BTC}'s sensitivity to electricity prices in \ref{sec:mining}.} However, some of these sources are carbon-intensive, raising environmental concerns\footnote{Cases such as Iran \cite{Nikzad2022Techno-economicMining}, Kazakhstan \cite{GhaebiPanah2022InvestmentMining,Coinshares2022TheImpact,Oysti2021BitcoinConsumption}, parts of China,\cite{Roeck2022LifePlant,Coinshares2022TheImpact,Carter2021BitcoinZero}, Venezuela \cite{GhaebiPanah2022InvestmentMining}, and parts of the \ac{US} and Canada \cite{Coinshares2022TheImpact,Oysti2021BitcoinConsumption}, where outdated natural gas power plants are retrofitted for bitcoin mining \cite{Roeck2022LifePlant}, often take center stage. Criticism is sometimes also directed at green energy sources, as their use for mining could allegedly reduce green energy availability for other purposes \cite{Amir2021RenewableWoes}.} This is particularly relevant given the strong advocacy for limiting global temperature increases to 1.5 degrees Celsius \cite{IPCC2018GlobalC}.


While some mining relies on carbon-intensive energy sources, the extent of this reliance is debated \cite{Paez2022RFITransition}. China's \cite{CCAF2022BitcoinResurgence,Oysti2021BitcoinConsumption,McCook2022DriversEmissions,Carter2021BitcoinZero} and Kazakhstan's \cite{Batten2023LeavingGreen} cryptocurrency bans have furthermore shifted the landscape,\footnote{See Coinshares \cite{Coinshares2022TheImpact} (also \cite{Carter2021BitcoinZero}) for the opposite argument that the China ban had a milder effect on Bitcoin decarbonization than commonly estimated, and that the primary effect was a reduction of seasonality in emissions.} consolidating the pre-existing trend to mine bitcoin in the \ac{US} and potentially influencing the industry's carbon footprint.

\subsection{Estimating Bitcoin's Carbon Footprint}
\label{sec:consumption}

A key question in the literature is the extent of Bitcoin's carbon footprint. Although Bitcoin's energy consumption is undeniably high compared to systems like \ac{PoS} \cite{OSTP2022ClimateStates,Ibanez2023TheExpansionb,Platt2021TheProof-of-Work,Read2022GreenwashingIndustry,GDF2022Re:Assets,Gallersdorfer2022EnergyProtocols}, disagreements persist regarding its precise magnitude, suitable metrics for representation, and the carbon intensity of its energy consumption.

\paragraph{Data sources}

Estimates of the magnitude of Bitcoin's energy consumption vary widely. The White House \ac{OSTP} \cite{OSTP2022ClimateStates} suggests that it ranges between 72 and 185 billion kWh per year.

The CCAF's (Cambridge Centre for Alternative Finance) CBECI (Cambridge Bitcoin Energy Consumption Index) is a widely used source.\footnote{https://ccaf.io/cbeci/index} For \ac{GHG} estimates, CBECI uses the carbon intensity of the grid mix corresponding to mining pools' IP addresses, overlooking \ac{BTM} renewable mining. In contrast, the \ac{BMC} prioritizes first-hand data from miners \cite{BMC2022Bitcoin2022}, which has its limitations due to self-reporting and data comparability (see causality theories discussion below). Non-scientific sources are also common in public discourse.\footnote{For instance, De Vries' Dogecoin blog ``Dogeconomist" (currently ``Digiconomist" \cite{DeVries2014DogeconomistDigiconomist}) features a \ac{BECI} frequently cited in news articles \cite{Carter2022CommentsMining}.

De Vries employs a ``top-down" approach, estimating the share of miners' revenue spent on electricity. This method is generally considered unrealistic, and a ``bottom-up" approach, which estimates energy consumption per hash based on the hash rate, is preferred \cite{CCRI2022TheNetwork,CCRI2022EnergyBlockchain}.}

\paragraph{Indicators, Metrics, and Attribution}
\label{paragraph:magnitudes}

Various indicators have been used to portray Bitcoin's environmental impact. One area of disagreement concerns appropriate comparisons:

\begin{description}
   \item \textbf{To countries}: For example, De Vries' \ac{BECI} compares Bitcoin's annual \ac{CO2}, electricity, and IT waste to countries like Switzerland, Colombia, and the Netherlands , respectively \cite{deVries2022BitcoinIndex} (see also \cite{OSTP2022ClimateStates}).
   
   \item \textbf{To industries:} The previous approach has been criticized for comparing ``apples to oranges," as many industries also surpass individual countries in energy consumption. Bitcoin reportedly consumes less energy than steel and aluminum \cite{Carter2021BitcoinZero}, gold \cite{Carter2021BitcoinZero}, the banking industry \cite{Oysti2021BitcoinConsumption}, the global monetary system \cite{Coinshares2022TheImpact}, Christmas lights \cite{Komando2012VampireBills}, idle electric appliances \cite{NavigantConsultingInc2008EnergyApplications}, and others\footnote{"Aviation Industry, Marine Transport Sector, Air Conditioners and Electric Fans, Data Centers, and Tumble Dryers" \cite[12]{Coinshares2022TheImpact} (see also \cite{Carter2021BitcoinZero,Velicky2023RenewableBitcoin}).} which yield more favorable comparisons for Bitcoin.
   
\end{description}

There is also disagreement about the best global denominator for representing Bitcoin's share:

\begin{description}
    \item \textbf{A share of global \textit{electricity} consumption}: For example, De Vries \cite{deVries2022BitcoinIndex} and Oysti \cite{Oysti2021BitcoinConsumption}.
    
   \item \textbf{A share of global \textit{energy} consumption}: Some argue that focusing only on electricity consumption obscures conversion efficiencies (\cite{Carter2021BitcoinZero,Paez2022RFITransition}, see also \cite{Oysti2021BitcoinConsumption}).\footnote{Natural gas, hydroelectric, coal, and nuclear generation have conversion factors of 44\%, 90\%, 32\%, and 32\%, respectively. As Bitcoin mining relies more on the first two than the average industry, its share of global energy consumption is lower than its share of global electricity consumption. In 2019, Bitcoin's electricity consumption would have placed it 28th in a 70-country world ranking, but 63rd in terms of \textit{energy} consumption. This would be even more so the case for carbon emissions if Bitcoin's energy mix is indeed greener than the average sector's, as advocates suggest \cite{McCook2022DriversEmissions}.}
   
    \item \textbf{A share of global \ac{CO2} \textit{emissions}}: High energy consumption may not necessarily lead to equally high emissions\cite{McCook2022DriversEmissions},\footnote{According to Coinshares \cite{Coinshares2022TheImpact}, bitcoin mining accounts for just 0.08\% of global carbon emissions (0.07\% for McCook \cite{McCook2022DriversEmissions}), compared to 0.14\% of global energy consumption and 0.50\% of electricity consumption.} and climate change is a function of the latter, not the former.
  
    \item \textbf{As a share of global \textit{\ac{GHG}} emissions}: \ac{CO2} emissions do not encompass the entirety of \ac{GHG} emissions \cite{McCook2022DriversEmissions,OSTP2022ClimateStates}.     
\end{description}

There is also controversy about the basis of comparison for contrasting Bitcoin's consumption/emissions with its value proposition. Alternatives include:

\begin{description}
   \item \textbf{A per-transaction basis}: This approach is taken by De Vries\footnote{See De Vries' statement that a bitcoin transaction requires a carbon footprint of 424.40 kg\ac{CO2}, and an electricity requirement of 760.90 kWh, equaling 940.616 Visa transactions, the average power consumption of a US household for 26.08 days, and the weight of 2.59 iPhones, respectively \cite{deVries2022BitcoinIndex}.} \cite{deVries2022BitcoinIndex} (see also \cite{OSTP2022ClimateStates,CCRI2022DeterminingNetworks}) and typically considers only \ac{L1} transactions, leading to criticism \cite{Paez2022RFITransition,Imran2018TheMining} (see per-dollar below). It has also been criticized because mining, not throughput, is the cause of Bitcoin's electricity consumption \cite{Ibanez2023TheExpansion,Ibanez2023DontBitcoin,Paez2022RFITransition}. Mining depends more on Bitcoin price than the number of transactions, which only has an indirect influence \cite{Paez2022RFITransition}. This may make this metric misleading, as it suggests that for Bitcoin's throughput to grow, it needs to consume even more energy, a statement that may hold for \ac{PoS} currencies but not for \ac{PoW} ones \cite{Ibanez2023TheExpansion,Ibanez2023DontBitcoin}.
   
   \item \textbf{A cumulative transactions basis}: Because the energy expenditures to mine a present bitcoin secure the entire history of \textit{past} transactions, and not just the coinage of the latest coin \cite{Imran2018TheMining}.\footnote{Imran argues that the key trend is not that energy consumption per transaction is growing, but that the total value secured is increasing \cite{Imran2018TheMining}.}
   
   \item \textbf{On a per-dollar or per-coin \textit{settled} basis}: Given that \ac{L2} solutions such as the Lightning Network (or even changing Bitcoin's parameters) may allow Bitcoin to scale arbitrarily without increasing energy usage \cite{Ibanez2023TheExpansion} this is often seen as a more adequate metric \cite{Paez2022RFITransition,Imran2018TheMining}.

\item \textbf{A per-dollar or per-coin \textit{mined} basis:} This approach, suggested by Jones et al \cite{Jones2022EconomicGold}, offers a novel angle in the short run, but cannot be meaningfully applied over time as it neglects Bitcoin's decreasing emission rate.\footnote{Using this metric forces the user to simultaneously argue that when the last bitcoin is mined, Bitcoin's emissions will be infinite, and that approximately 90\% of Bitcoin's climate damages have already occurred and the rest will be spread over an increasingly carbon-neutral energy grid.} It also assumes an "origin accounting" methodology (see below).
\end{description}

One may also identify different carbon accounting approaches used in the field of Bitcoin \cite{WBD2022CanCross,Ibanez2023DontBitcoin,Gallersdorfer2021AccountingSystems,SouthPole2022AccountingImpacts}:

\begin{description}
   \item \textbf{Transaction accounting:} Determines the total carbon footprint of a block and divides the total by the number of transactions contained in the block. See \ref{sec:bitcoin}.
   
   \item \textbf{Origin accounting:} Conducts a genealogical analysis to consider the carbon emissions historically necessary to produce each block.
   
   \item \textbf{Maintenance accounting:} Attributes the carbon footprint to the \textit{holding} of a coin, as the demand for the coin ultimately incentivizes mining (see also \cite{Coinshares2022TheImpact}).
   
    \item \textbf{Hybrid accounting:} Combines transaction accounting, applied to emissions derived from the pursuit of transaction fees, with maintenance accounting, applied to emissions derived from the pursuit of block rewards \cite{Ibanez2023DontBitcoin,Gallersdorfer2021AccountingSystems}:
   
\end{description}

Moreover, various theories of causality are applied to Bitcoin's impact on the energy grid and the environment \cite{Stoll2018TheSeries,Ibanez2023DontBitcoin}:

\begin{description}
   \item \textbf{Marginal accounting:}     
   The \ac{BMC} is criticized for accepting miners' claims of high renewability when they are merely located in highly renewable areas and claiming the average grid mix. If the additional energy demand following the miners' installation is met with non-\ac{RE} sources, it is argued that the \textit{marginal} consumption cannot be considered green.

   \item \textbf{Attributional accounting:} The alternative viewpoint posits that there is no reason to prioritize older energy consumers over younger ones and that all consumers are on equal footing from a causal point of view concerning causing the need for electricity generation. Under this perspective, one may attribute energy consumption with simple averages or based on the \ac{EACs} such as \ac{RECs} or \ac{GOs} purchased.
\end{description}

Whichever theory of causality is preferred, it is crucial to apply it consistently. For example, if it is not legitimate to claim the average grid mix when it is highly renewable, it is hard to argue for the legitimacy of most journal articles, press, and activism that also take mining pools' IP address and attribute the average emissions of the corresponding area.

The upcoming introduction of carbon accounting requirements \cite{SEC222SECInvestors} may provide additional insight into these debates.

\begin{figure*}
\centering
\begin{displaymath}
\text{Controversies}
\begin{cases}
\text{Green mix estimation} &
\begin{cases}
\text{Grid mix attribution} &
\begin{cases}
\text{Bottom-up}\\
\text{Top-down}
\end{cases} \\
\text{Self-reported attribution}
\end{cases} \\
\text{Causal theories} &
\begin{cases}
\text{Origin accounting}\\
\text{Transaction accounting}\\
\text{Maintenance accounting}\\
\text{Hybrid accounting}\\
\end{cases}\\
\text{Emission factors} &
\begin{cases}
\text{Marginal accounting}\\
\text{Attributional accounting} &
    \begin{cases}
    \text{Average attribution}\\
    \text{\ac{EACs}-based attribution}
\end{cases}\end{cases}\\
\text{Metrics} &
\begin{cases}
\text{Denominators} &
\begin{cases}
\text{Transactions} &
\begin{cases}
\text{Current}\\
\text{Cumulative}
\end{cases} \\
\text{Value} &
\begin{cases}
\text{Settled} &
\begin{cases}
\text{Fiat}\\
\text{\ac{BTC}}
\end{cases}\\
\text{Mined} &
\begin{cases}
\text{Fiat}\\
\text{\ac{BTC}}
\end{cases}
\end{cases}\\
\text{Global consumption} &
\begin{cases}
\text{of electricity} \\
\text{of energy}
\end{cases}\\
\text{Global emissions} &
\begin{cases}
\text{of \ac{CO2}}\\
\text{of \ac{GHG}}\\
\end{cases}\\
\end{cases}
\end{cases}
\\
\text{Comparison reference} &
\begin{cases}
\text{Countries} \\
\text{Industries}
\end{cases} \\
\end{cases}
\end{displaymath}
\caption{The controversy about Bitcoin mining's environmental impact.}
\label{diagram:controversy}
\end{figure*}

\paragraph{Controversies over Bitcoin's future environmental impact}

A significant area of debate is the projection of Bitcoin's future energy consumption \cite{Carter2021BitcoinZero,OSTP2022ClimateStates}. Critics express concern about the potentially increasing energy requirements as Bitcoin becomes more mainstream, while advocates argue that Bitcoin leads to a higher standard of living, which in turn may result in lower emissions through the environmental Kuznets curve \cite{Carter2021BitcoinZero}.

Advocates also highlight that critics often overlook the effect of \textit{halvings} in their projections \cite{Carter2021BitcoinZero}. Every 210,000 blocks (approximately 4 years), Bitcoin mining block rewards are cut in half, a process that is expected to continue until 2140, when the last bitcoin will be mined \cite{McCook2022DriversEmissions,Oysti2021BitcoinConsumption}. As the incentive to mine decreases, energy consumption may reduce as well. Carter and Stevens \cite{Carter2021BitcoinZero} (see also \cite{Paez2022RFITransition}) expect Bitcoin mining emissions to peak at 1\% of global emissions at worst.\footnote{If they have not peaked already due to ``the mass migration away from worst-in-class Chinese coal, to best-in-class (or at least 50th-percentile) Natural Gas, an emissions drop in emissions of between 70 to 80\% per unit of energy, despite energy use trending upwards. In other words, for emissions to return to pre-China migration levels, energy expenditure would need to grow three-fold, and the demonstrably false assumption that there will never be any further efficiency gains in mining hardware" \cite{McCook2022DriversEmissions}.}

It is important to note three related observations. First, peaking as a percentage of global emissions or energy/electricity consumption is not the same as peaking in absolute terms. Bitcoin mining's consumption may decline as a share of global consumption while still increasing in absolute numbers, especially if global energy consumption continues to grow. Second, the expected threefold electrification of the world in the next decades may lead to a larger increase in global \textit{electricity} consumption than in global \textit{energy} consumption, causing Bitcoin's electricity consumption to peak significantly earlier as a percentage of global energy consumption \cite{Carter2021BitcoinZero}. Third, if Bitcoin's electricity consumption peak arrives markedly earlier than the world's electrification peak, this would result in more \ac{GHG} emissions than otherwise.

\paragraph{Other impacts}

Bitcoin mining has been criticized for the electronic waste (\textit{e-waste}) it might produce \cite{Mankala2022AnImpact,deVries2021TheUse}. De Vries and Stoll \cite{deVries2021BitcoinsProblem} provide the most pessimistic estimate of 30.7 metric kilotons (0.07\% of all e-waste) \cite{Paez2022RFITransition}. They argue that as more-efficient \ac{ASIC}s enter the zero-sum mining market and drive up the hash rate, older \ac{ASIC}s are pushed out due to increasing costs and an upward break-even threshold \cite{Coinshares2022TheImpact,Mellerud2021BitcoinTexas,Yazc2023AHardware}. However, De Vries and Stoll's assumptions have been criticized (\cite{Saylor2022BitcoinAgency}), with arguments highlighting the non-zero price at which supposedly obsolete \ac{ASIC}s trade, the recyclability of \ac{ASIC} components, and how, under a buyer of last resort model (see \ref{subsec:models}), old miners do not become e-waste as assumed but rather serve as ancillary infrastructure for peak supply events or are recycled \cite{Saylor2022BitcoinAgency}.

Another environmental concern related to Bitcoin is noise pollution \cite{NguyenAnhHoang2018ReusingHouse}. Although there is little controversy surrounding this issue \cite{OSTP2022ClimateStates}, it is worth noting that noise pollution is a byproduct of all data centers and not specific to Bitcoin \cite{Voell2022WeProblem}.

\paragraph{The underlying philosophical debate}

The debate surrounding Bitcoin's environmental impact is closely connected to the broader controversy about whether Bitcoin has intrinsic value \cite{Paez2022RFITransition,Rennie2021ClimateBitcoin,Coinshares2022TheImpact,Carter2021BitcoinZero}. Advocates of Bitcoin argue that its detractors consider its environmental impact to be excessive primarily because they assume its lack of value. However, if critics recognized the services that Bitcoin provides, they may not reach this conclusion \cite{Carter2021BitcoinZero, Paez2022RFITransition,Rennie2021ClimateBitcoin}.

This is closely linked to the question of whether \ac{PoW} itself has any value compared to alternative consensus mechanisms like \ac{PoS} \cite{Paez2022RFITransition,Coinshares2022TheImpact}. It is important to note that while \ac{PoW} is energy-intensive by design, it is not the sole cause of Bitcoin's high energy consumption. Other factors, such as small block size and large block time, also concur with \ac{PoW} to cause its high energy use. Although many in the Bitcoin community support these features, alternative forks of Bitcoin have emerged with lower energy consumption by modifying these parameters \cite{Lin2020BlockchainCryptos}. Regulators should consider this when evaluating potential \ac{PoW} bans.

\subsection{Regulatory approaches}
\label{sec:regulatory}

Decision-making on laws and directives can prohibit, constrain, or promote the usage of a public or private good in various policy fields. A similar case applies to digital and distributed ledger technologies. Governments are currently adopting differing approaches, ranging from restrictive to supportive.

China initiated a ban on Bitcoin-related activities, such as exchange and financing between fiat money or coin substitutions, in 2017 in order to maintain regulatory consistency \cite{Xie2019WhyChina}. In September 2019, the country extended the ban to all crypto assets, including mining activities \cite{Prasad2021ChinaBitcoin}. The primary reasons cited were combating financial crimes, preventing capital flight, and maintaining economic stability through increased state intervention \cite{Shin2022WhatsBan}.

In contrast, the \ac{US} has adopted a more supportive approach. The Executive Order on Ensuring Responsible Development of Digital Assets was signed in March 2022, recognizing the importance of \ac{DLT}s while calling for higher standards to mitigate various risks, such as financial threats, consumer and investor protection, and climate and environmental impacts \cite{Biden2022ExecutiveAssets}. This led to further research, including the Report on Climate and Energy Implications of Crypto-Assets in the \ac{US} \cite[3]{OSTP2022ClimateStates}, which emphasized the potential for using clean energy in mining to reduce \ac{GHG} emissions.

The European Union's approach is more complex. A uniform regulatory framework for all crypto-assets has been introduced across the Union to improve consumer protection, increase market integrity and financial stability, and prevent criminal activities such as market manipulation, money laundering, and terrorist financing \cite{EuropeanParliament2022CryptocurrenciesThreats}. However, the \ac{EU} Commission has criticized \ac{PoW} mining for its high electricity consumption, particularly in light of energy shortages following Russia's reduced gas supply. Potential measures include an energy efficiency label for \ac{PoW} mining and a grading mechanism to encourage a shift towards systems perceived to be more environmentally friendly, such as \ac{PoS} \cite{Ainger2022EUPlan,Shawn2022EULaw}.

\section{Limitations of \ac{RE}}
\label{sec:limitations}

Various obstacles hinder the widespread adoption of \ac{RES} necessary for a renewable-powered world. The primary challenges are profitability \cite{Carter2021BitcoinZero} and intermittency \cite{Guo2018OptimizationResource,Brook2014WhyMix,SolunaComputing2022Re:Assets}. \ac{RE} generation has not traditionally been cost-effective for mass adoption by producers \cite{Carter2021BitcoinZero}. Governments have relied on subsidies and similar strategies to compensate for this, but such measures are costly and come with their own problems. While \ac{RE} generation is becoming more efficient and less expensive. With this trend expected to continue, high levels of renewable penetration may present challenges that impact profitability \cite{Carter2021BitcoinZero}.

\subsection{Imbalances}
\label{sec:imbalances}

For \ac{VRE} generation, the most well-known manifestation of these challenges is the so-called \textit{duck curve} \cite{Quirk2021CryptocurrencyInnovation.,Oysti2021BitcoinConsumption,BCEI2021BitcoinFuture}. This phenomenon arises from the fact that sunlight is available only during the day and wind, though more unpredictable, typically blows more strongly at night. Meanwhile, energy demand peaks in the late afternoon and early evening when people use appliances upon returning home \cite{BCEI2021BitcoinFuture}.

This and other forms of renewable volatility, along with other sources of intermittency (e.g., transmission constraints and extreme weather events), lead to imbalances that may result in negative pricing or \textit{curtailment} at high levels of renewable penetration \cite{Paez2022RFITransition,McCook2022DriversEmissions,Shan2019BitcoinCaiso,BCEI2021BitcoinFuture,GhaebiPanah2022InvestmentMining,Dogan2022AreMethod,Brook2014WhyMix,Frumkin2021EconomicsEnergy,Joos2018Short-termGermany}. This issue is particularly unique to electricity, as it must be consumed almost immediately after being produced \cite{Mellerud2021BitcoinTexas}.

The fundamental problem of imbalance\footnote{Technically, the grid operator's very role is to prevent imbalances by matching supply and demand, e.g., by curtailing supply. However, in this scenario one may speak of an imbalance in a broader sense. Unavoidable deviations from contracted positions result from imperfect forecasts and unforeseen events (outages of power plants and lines), which the system operator corrects by calling upon reserve capacity that has been contracted. Although the cost of this capacity is partially recovered through grid fees and penalties for deviating from schedule, around two-thirds of the costs are socialized in the form of higher prices \cite{Joos2018Short-termGermany}.} occurs between exogenously fluctuating \ac{VRE} sources and variable electricity demand throughout the day, week, or year. A well-functioning energy system needs \textit{load following} to ensure the network never overloads nor experiences blackouts \cite{Carter2021BitcoinZero,Sarquella2022BitcoinAccelerator}.

However, \ac{VRE} generation typically cannot be increased at will (and can only be decreased by curtailing, which entails significant opportunity costs) due to its inherent intermittency \cite{Oysti2021BitcoinConsumption,Carter2021BitcoinZero}. Although supply and demand sometimes align, mismatches often occur \cite{Rennie2021ClimateBitcoin,Carter2021BitcoinZero}. These fluctuations can be sudden, exacerbating the challenges \cite{MellerUd2021BitcoinTexas}. Demand-side fluctuations may also be sudden, such as during abrupt heat or cold waves \cite{Mellerud2021BitcoinTexas}.

These discrepancies are problematic as they affect grid resiliency. In the absence of alternative solutions, preserving resiliency requires capping the contribution of \ac{VRE} to the grid at a relatively low percentage and supplementing peak load with non-renewable sources. This constraint impedes the progress of decarbonization. Some partial solutions to these issues are discussed in Table \ref{table:mitigation}.


\begin{table*}[ht]
  \centering
  \begin{tabular}{|p{2cm}|p{5.5cm}|p{6cm}|}
    \hline
    \textbf{Strategy} & \textbf{Description} & \textbf{Challenges} \\ \hline
    \textbf{Transmission} & Importing and exporting energy from areas with excess supply to areas with excess demand is an effective way to balance electricity markets \cite{Jenkins2022ElectricityAct,Mellerud2021BitcoinTexas,BCEI2021BitcoinFuture}. &  Transmission lines have limited capacity, experience congestion, struggle to keep up with electrification trends, suffer from energy losses proportional to their length, and require substantial initial investments \cite{BCEI2021BitcoinFuture,Rennie2021ClimateBitcoin,Carter2021BitcoinZero,Menati2022ModelingStudy,Coinshares2022TheImpact}. "Stranded" energy cannot be transmitted. \ac{RE} generation is often most efficient in remote locations. \\ \hline
    \textbf{Capacity Expansion} & Investing in excess \ac{RE} infrastructure to meet demand during low supply periods. & Over-building or over-investing impacts the sector's profitability, leading to low or negative prices during high supply periods and necessitating government subsidies \cite{Quirk2021CryptocurrencyInnovation.,Mellerud2021BitcoinTexas,BCEI2021BitcoinFuture} and curtailment, often intentionally designed in capacity expansion projects \cite{Frew2021TheSystems}. \\ \hline
    \textbf{Curtailment} & When \ac{RE} production infrastructure is built, excess energy is wasted to avoid issues such as overloading transmission capacity or negative pricing \cite{Shan2019BitcoinCaiso,Bird2014WindStates,Paez2022RFITransition,McCook2022DriversEmissions,SolunaComputing2022Re:Assets}. & Curtailment has an opportunity cost in terms of unsold energy, decreasing the profitability of \ac{VRE} generation. Curtailment is projected to increase over time \cite{Braiins2021OptimizationsSources,Joos2018Short-termGermany}. \\ \hline
    \textbf{Storage} & Storing energy during excess supply and using it during excess demand periods through batteries (see \ref{sec:alternatives} or other methods like pumped hydroelectric storage \cite{Mellerud2021BitcoinTexas,Eid2021EnhancedLoad,BCEI2021BitcoinFuture,Velicky2023RenewableBitcoin}. & Batteries and storage solutions are expensive \cite{Mellerud2021BitcoinTexas,Eid2021EnhancedLoad,Carter2021BitcoinZero} and have limited capacity, restricting their large-scale use (see \ref{subsection:Batteries}). \\ \hline
    \textbf{Demand-Response Programs} & Grid operators influence electricity demand patterns to match supply patterns, using \ac{FLR} (foreseenin voluntary \ac{PPA}s or \ac{DR} programs) to compensate energy customers for not consuming energy during peak events \cite{Guo2018OptimizationResource,BCEI2021BitcoinFuture,Paez2022RFITransition,Mellerud2021BitcoinTexas,Menati2022ModelingStudy}. & Most loads are not flexible enough for large-scale implementation without significant costs or opportunity costs. \\ \hline
  \end{tabular}
  \caption{An overview of some of the main strategies to counter problems of imbalance spurred by \ac{VRE}.}
  \label{table:mitigation}
\end{table*}

Note that \ac{DR} programs may be regarded as a form of sector coupling (see \cite{Ramsebner2021TheReview}). Furthermore, storage and \ac{DR} programs are subtypes of so-called "power-to-X" solutions, i.e. the practice of converting or storing surplus electric power during excess supply periods \cite{Sternberg2015Power-to-WhatSystems,KCE2016EnergyChemistry,Lund2015ReviewElectricity,GhaebiPanah2022InvestmentMining}.

\subsection{Funding Challenges}
\label{sec:funding}

\ac{RE} projects face various financial challenges. For instance, ``solar value deflation" makes it difficult to attract investors and developers to build solar plants, despite the necessity of these investments for the energy transition (Irena in \cite[6]{Paez2022RFITransition}). 

The construction of \ac{RE} generation facilities is often delayed or not completed, and their remote locations necessitate additional transmission investment costs and increase transmission losses, reducing profitability \cite{Bastian-Pinto2021HedgingMining,Paez2022RFITransition,Mellerud2021BitcoinTexas,Coinshares2022TheImpact,Carter2021BitcoinZero}. \ac{RE} projects also encounter connection queues due to technical or regulatory reasons \cite{BCEI2021BitcoinFuture,Rand2021Queued2020,Bastian-Pinto2021HedgingMining,Paez2022RFITransition}. \footnote{Rand et al \cite{Rand2021Queued2020} find that the average US commercial power projects spent approximately 3.5 years waiting for connection approval between 2010 and 2020, with connection wait times on an upward trend and ~680 GW of zero-carbon generation capacity stuck in the queue.}

Spot price volatility in energy markets, exacerbated by these connection queues, poses additional challenges. Sellers waiting to sell at a regulated price must either sell energy in the spot market at the risk of price volatility or potentially miss their contracted connection date by not coming onstream to the grid earlier \cite{Bastian-Pinto2021HedgingMining}.

\subsection{Prospects}
\label{sec:prospects}

Despite the challenges, the practices and technologies discussed above are essential to meeting various \ac{IEA} goals for decarbonization, \ac{RE} penetration, grid resiliency, and electrification \cite{IEA2020Renewable2021,IEA2021NetSector}. \footnote{While not entirely carbon-free, \ac{RE} generation has significantly lower life-cycle emissions compared to fossil generation on average \cite{WNA2022CarbonElectricity,Frumkin2021EconomicsEnergy}.}

In this context, \ac{PoW} mining emerges as an alternative that can provide additional income and ancillary services (auxiliary services designed to provide stability to the energy grid), including reactive power and voltage control, frequency control, scheduling and dispatch of contingency energy supply reserves, flexibility energy supply reserves for outages, and flexible energy demand load \cite{Rhodes2021ImpactsERCOT,Mellerud2021BitcoinTexas,Paez2022RFITransition}: ``power-to-bitcoin."

\section{Distinctive Characteristics of \ac{PoW} Mining}
\label{sec:mining}

As an economic activity, \ac{PoW} mining in general and Bitcoin mining in particular exhibit several unique characteristics that set it apart as an energy buyer \cite{BCEI2021BitcoinFuture}. These distinctive features include:

\paragraph*{Flexibility of load}

Bitcoin miners can be activated or deactivated rapidly, with sub-second responsiveness \cite{Quirk2021CryptocurrencyInnovation.,Oysti2021BitcoinConsumption,BCEI2021BitcoinFuture,Paez2022RFITransition,Mellerud2021BitcoinTexas}. \footnote{Lancium \cite{Rhodes2021ImpactsERCOT} in Mellerud \cite[45]{Mellerud2021BitcoinTexas} states that the reaction time is 5 seconds, comparable to the fastest-reacting peaking plants.} This capability allows for high load flexibility without the constraints of inertia, cooling,\footnote{Cooling costs exist but do not increase (and may even decrease) with flexibility of load \cite{Mellerud2021BitcoinTexas}.} warming up requirements, or other reaction costs \cite{Mellerud2021BitcoinTexas}. As a result, Bitcoin mining may contribute to grid stability \cite{Paez2022RFITransition}.

A key factor for the flexibility of load is the \textit{availability} of load. A load resource is ``available" if it is continuously demanding energy, and hence can be turned off. As many modern \ac{ASIC}s typically operate at full capacity or even overclock, they provide a reliable and stable load with considerable availability \cite[45]{Mellerud2021BitcoinTexas}. To maintain grid resiliency, availability must be consistent over the long term. Bitcoin mining meets this requirement due to miners' extended time horizons\footnote{Influenced by ideological and economic factors \cite{Bonaparte2022TimeSpeculative,Ammous2022HardPreference} and the need for long-term planning when entering into \ac{PPA}s \cite{Mellerud2021BitcoinTexas}.} and the fact that the cash-flow break-even level is generally much lower than the \ac{ROI} break-even level \cite{Mellerud2021BitcoinTexas}.\footnote{However, volatility can negatively impact the stability of availability (see \ref{subsec:difficulties}).}

\paragraph*{Interruptibility}

The Bitcoin protocol is designed to produce a new block every 10 minutes on average. If the mean block time deviates from this value, the ``difficulty level" is adjusted to restore the average interval to 10 minutes. Save small deviations in between difficulty level adjustments, the expected block time remains \textit{always} 10 minutes,\footnote{If 9 minutes have passed since the last block was mined and no new block has yet been found, the expected block time is still 10 minutes, not one minute \cite{McCook2022DriversEmissions}.} regardless of the time elapsed since the last block was mined \cite{McCook2022DriversEmissions}. Since mining involves guessing the nonce by attempting millions of terahashes per second, the vast majority of which are incorrect, interrupting the mining process results in no lost work.

In other words, mining relies on \textit{non-time-sensitive computation} \cite{Mellerud2021BitcoinTexas}, which, combined with rapid reaction times \cite{Mellerud2021BitcoinTexas,Hajipour2022AnMicrogrids}, allows for immediate output switching \cite{Bastian-Pinto2021HedgingMining,OSTP2022ClimateStates}. The quick response time and near-zero reaction costs (no output loss)\footnote{Freier and Ibañez \cite{Freier2023BitcoinGermany} suggest that interruptions to mining may incur economic costs at very small scales, as connecting to mining pools for income stability often results in reduced rewards for miners providing hash rate in an interrupted manner.} result in high interruptibility \cite{Oysti2021BitcoinConsumption,Mellerud2021BitcoinTexas,Guo2018OptimizationResource,Paez2022RFITransition,BCEI2021BitcoinFuture}. These distinctive characteristics make \ac{PoW} mining a unique energy buyer that can support grid stability and contribute positively to the energy sector.

\paragraph*{Portability/mobility}

Bitcoin mining is location-agnostic \cite{Oysti2021BitcoinConsumption, BCEI2021BitcoinFuture}, as it requires minimal investment in immovable assets. \ac{ASIC}s and other mining equipment are easily transportable, as are most supplementary hardware components. Mining can operate in diverse geographies and climates without grid connections, only requiring an electricity source \cite{McCook2022DriversEmissions, Paez2022RFITransition, Rennie2021ClimateBitcoin}, in contrast to most other industries \cite{Mellerud2021BitcoinTexas}. The portability of Bitcoin mining is further enhanced by modularised solutions, such as shipping-container-based systems that allow for mining operations anywhere on Earth \cite{McCook2022DriversEmissions, Rennie2021ClimateBitcoin, Carter2021BitcoinZero}. Mobility also increases the resiliency of the Bitcoin network \cite{Rennie2021ClimateBitcoin}.

Although there are some immovable investments, such as land and rack space\footnote{Quirk \cite{Quirk2021CryptocurrencyInnovation.} (see also \cite{Velicky2023RenewableBitcoin}) notes that rack space requirements are reduced with liquid immersion cooling, enabling more dispersed, cellular mining facilities in remote locations.}, the speed at which mining migrated to other locations after the China ban on mining \cite{McCook2022DriversEmissions, Velicky2023RenewableBitcoin} and the seasonal nature of mining operations in China \cite{Carter2021BitcoinZero, Velicky2023RenewableBitcoin} empirically demonstrate the significant geographic flexibility of mining \cite{Mellerud2021BitcoinTexas}.

Additionally, mined bitcoin only require an internet connection for transfer, eliminating the need for transportation infrastructure like pipelines, trains, trucks, or flights. The only transportation needed is for shipping \ac{ASIC}s and cooling equipment once \cite{Mellerud2021BitcoinTexas}.\footnote{Note that, "in the first years of \ac{ASIC}s mining, miners were significantly less geographically flexible than they are now, since they had such big advantages of locating themselves in China, close to their suppliers of \ac{ASIC}s," but this is no longer the case, as ``the technological improvement rate of \ac{ASIC}s has drastically slowed down, leading to a longer life-time and thus less importance of having the newest gear." \cite[48-49]{Mellerud2021BitcoinTexas} (see also \cite{Imran2018TheMining})} This enables high responsiveness to market conditions and enhances the provision of ancillary services, which need to be located near production areas \cite{Menati2022ModelingStudy, Strohle2016LocalGrids, Wang2021AMethod, Walton2022BitcoinComplicated., Cohn2019IsResponse, King2022BitcoinBehind-The-Meter}. Bitcoin mining is also not labor-intensive, allowing the placement of mining farms far from urban centers.

\paragraph*{Price Sensitivity}

Bitcoin mining is one of the most price-sensitive industries in the world \cite{Gronowska2021GreenHeat}, with few inputs (mainly electricity) and outputs (primarily \ac{BTC}). As \ac{BTC} prices are ``invariant across times and locations" \cite{Paez2022RFITransition}, this results in cost sensitivity (specifically OPEX sensitivity), and electricity becomes a crucial cost factor \cite{Mellerud2021BitcoinTexas}, driving the intense competition for affordable energy in the Bitcoin market. Furthermore, cost sensitivity is expected to increase in the coming years (see \ref{subsec:second}).

As a result, Bitcoin \ac{ASIC}s are better suited to react to volatile electricity costs and are more compatible with placement on the generation side than conventional assets \cite{Shan2019BitcoinCaiso}. Moreover, the different profitability profiles of various \ac{ASIC}s offer significant complementarities with multiple energy system niches and patterns \cite{Paez2022RFITransition} (granularity \cite{Mellerud2021BitcoinTexas}). Additionally, more efficient behavior is facilitated because break-even points for different actors are \textit{known} \cite{Paez2022RFITransition}.

\paragraph*{Scale agnosticity}
Bitcoin mining operations exhibit a high degree of scalability and adaptability, ranging from small-scale home mining\footnote{Note however that a minimum scale is required to participate in formal \ac{DR} programs. Mellerud places this in 100 kilowatts, which approximates a mining farm comprised of 70 Antminer S9 \cite{Mellerud2021BitcoinTexas}.} to large-scale industrial operations consuming gigawatts of power \cite[5]{Paez2022RFITransition}\cite{Roeck2022LifePlant}. This adaptability stems from the energy intensity of Bitcoin mining \cite{Mellerud2021BitcoinTexas}, enabling the industry to act as a shock absorber for the electrical grid \cite{Oysti2021BitcoinConsumption}.

\paragraph*{Consumption-level granularity}

The flexibility provided by energy-intensive \ac{ASIC}s with varying break-even points allows for rapid and precise adjustments in energy consumption levels, without incurring extra costs \cite{Mellerud2021BitcoinTexas,Hajipour2022AnMicrogrids}). This is in contrast to binary consumption options, where systems either consume at full capacity or cease consumption entirely, referred to as \ac{CLR} and \ac{NCLR}, respectively \cite{Rhodes2021ImpactsERCOT,Mellerud2021BitcoinTexas,Braiins2021OptimizationsSources}\footnote{Mellerud notes that, in Texas, the \ac{NCLR} status is harder to obtain from regulators but offers additional benefits to market participants \cite{Mellerud2021BitcoinTexas}.}.

\paragraph*{Non-rival energy consumption}

When considering energy consumption levels, it is essential to differentiate between the \textit{amount} of energy consumed and the \textit{type} of energy utilized. Mining's energy consumption may not necessarily lead to increased energy generation or emissions, as mining can utilize\footnote{And often does utilize, having led to a reduction of curtailing in Yunnan (China) in 2016 \cite{McCook2022DriversEmissions}.} otherwise curtailed energy \cite{McCook2022DriversEmissions,SolunaComputing2022Re:Assets}, stranded oil and gas \cite{Oysti2021BitcoinConsumption}, or flared gas \cite{Quirk2021CryptocurrencyInnovation.,Oysti2021BitcoinConsumption} (see \ref{sec:applications}). This indicates that miners may not be in direct competition with other energy consumers, and they do not always generate additional emissions. Instead, they can consume already-generated energy or act based on emissions that would have been produced regardless \cite{McCook2022DriversEmissions}.

\paragraph*{Diversification}

Bitcoin mining offers a valuable source of income diversification and stability for \ac{RE} sellers. The distinct and uncorrelated stochastic processes of global Bitcoin prices/hash rates and electricity prices enhance the value of switching outputs \cite[2]{Bastian-Pinto2021HedgingMining} (see also \cite{Sarquella2022BitcoinAccelerator}).

\paragraph*{Waste heat utilization}

Due to the law of conservation of energy, the energy input required to produce hashes results in a tangible output: heat \cite{NguyenAnhHoang2018ReusingHouse,Yazc2023AHardware}. The mining sector is currently developing cooling solutions to manage heat loads \cite{Quirk2021CryptocurrencyInnovation.,Gronowska2021GreenHeat}. Nevertheless, there are numerous innovative ways to repurpose this waste heat for other applications, such as residential heating (e.g., hot water and space heating for households and schools) and commercial use (e.g., greenhouses) \cite{NguyenAnhHoang2018ReusingHouse,Enachescu2019ClosedAlberta,Dogan2022AreMethod,Paez2022RFITransition}.

\begin{figure*}
\centering
\begin{tabular}{p{4cm}p{4cm}p{5cm}}
\hline
\multicolumn{3}{c}{\textbf{Salient Characteristics of Bitcoin Mining}} \\
\hline
\textbf{Category} & \textbf{Characteristic} & \textbf{Sub-characteristics} \\
\hline
Flexibility of Load & Availability of Load & Stability of Load \\
 &  & Reliability of Load \\
 &  & Long Time Horizon \\
\cline{3-3}
 & \multirow{2}{*}{Interruptibility} & \multirow{2}{*}{Quick Reaction Time} \\
\cline{1-1}Consumption Granularity & & \\
\cline{3-3}
 & Price Sensitivity & Bitcoin Price Sensitivity \\
 &  & Cost Sensitivity \\
 &  & Granularity \\
 &  & Information Completeness \\
 &  & Near-Zero Reaction Costs \\
\hline
Scalability & Scale Agnosticity & Scalability \\
 &  & Energy Intensity \\
\hline
Portability & Location Agnosticity & Movable Goods \\
 &  & Geography Independence \\
 &  & Modularized Solutions \\
 &  & Unnecessary Grid Connection \\
 &  & Low Labor Intensity \\
 &  & Transferability of Output \\
\hline
Other Characteristics & Non-Rivalrousness & \\
 & Non-Correlation & \\
 & Heat Output & \\
\hline
\end{tabular}
\caption{Salient characteristics of Bitcoin mining identified.}
\label{diagram:miningcharacteristics}
\end{figure*}

\section{Applications}
\label{sec:applications}

Due to the unique characteristics discussed in \ref{sec:mining}, Bitcoin mining can offer various services to the energy sector, including ancillary services, consumption of stranded resources, prevention of gas flaring, and provision of additional funding.

\subsection{Sectors}

\paragraph*{Wind energy}
Wind energy is impacted by the problem of imbalances. By employing \ac{FLR}, Bitcoin mining can provide shock-absorbing ancillary services \cite{Oysti2021BitcoinConsumption,King2022BitcoinBehind-The-Meter}. This can increase profitability by offering an alternative to selling energy at extremely low prices during periods of excess supply \cite{Oysti2021BitcoinConsumption,Hajipour2022AnMicrogrids}). This mining approach is typically conducted on-grid but \ac{BTM} (next to generation), which does not require additional transport infrastructure.

\paragraph*{Solar energy}
Solar energy also faces imbalances. \ac{PoW} mining can provide ancillary services in the form of \ac{FLR} for solar energy as well \cite{Hajipour2022AnMicrogrids,King2022BitcoinBehind-The-Meter}. Cogeneration systems based on solar generation and cryptocurrency mining have been shown to significantly increase the profitability of such enterprises \cite{Nikzad2022Techno-economicMining,Eid2021EnhancedLoad,King2022BitcoinBehind-The-Meter,Winton2021SolarPower,ARKInvest2021SolarBatteryBitcoin}\footnote{Eidt et al \cite{Eid2021EnhancedLoad} also find additional financial benefits of solar energy-bitcoin cogeneration unrelated to \ac{FLR} (see also \cite{Chehrehsaz2022SolarProfit}).}. This type of mining is generally performed on-grid but \ac{BTM}.

\paragraph*{Nuclear energy}
While nuclear energy supply patterns are more similar to oil and gas than wind and solar, there are constraints to adjusting generation levels to match demand patterns \cite{Denholm2012DecarbonizingStorage}. Technical and economic factors make operating a nuclear power plant in a load-following manner less cost-effective, primarily due to reactor cooling costs and reactor efficiency, which affect the variable cost of selling nuclear energy \cite{Denholm2012DecarbonizingStorage}. As a result, during periods of excess energy, the minimum price at which an energy seller is willing to sell its output may lead to negative prices, jeopardizing the sustainability of the business model. \ac{PoW} mining can also offer ancillary services in the form of \ac{FLR} in these situations, when mining is more efficient than operating the power plant in a load-following manner (for nuclear-based bitcoin mining with a stable load, see \cite{Yuksel2022BitcoinEnergy,Gonzalez2021InvestmentStates}). The use of nuclear reactors as power sources for data centers has already been proposed \cite{Howard2022Nuclear-poweredHorizon}.

\paragraph*{Waste gas recovery: venting and flaring}

Landfill gas and stranded natural gas are two significant sources of methane emissions to the atmosphere\footnote{Other initiatives also aim to utilize otherwise gas \cite{Rennie2021ClimateBitcoin}.}. Landfill gas consists of around 50\% methane \cite{USEPA2022BasicGas} and is often flared \cite{Decker2021BitcoinField}, which produces mainly methane due to inefficient combustion\footnote{``Flared natural gas burns the methane producing as a byproduct \ac{CO2}. This reduces theoretically the \ac{CO2} equivalents, but the efficiency of flaring varies largely, in some cases up to 70 percent can escape into the atmosphere" \cite[41]{Oysti2021BitcoinConsumption}. This also applies to other \ac{VOCs} \cite{Jacobs2020InnovatorsPower}.} and strong winds \cite{McCook2022DriversEmissions}.

Stranded natural gas is excess gas found near oil wells or exceeding a gas well's transmission capacity, which is unprofitable or impossible to process or transport for consumption elsewhere \cite{Quirk2021CryptocurrencyInnovation.,McCook2022DriversEmissions,Carter2021BitcoinZero,Decker2021BitcoinField}. This gas is also commonly flared \cite{Snytnikov2022FlareCapture}. If not flared, stranded and landfill gas is vented, resulting in even higher methane emissions \cite{Oysti2021BitcoinConsumption,Read2022GreenwashingIndustry,WB2022TurningWright}\footnote{Wright \cite{WB2022TurningWright} estimates that 70\% of US landfills freely emit methane into the atmosphere because they are too far from cities for methane to be processed (burned or refined into natural gas).}.

Considering the portability feature mentioned earlier, containerized mining and generator solutions \cite{McCook2022DriversEmissions} are being implemented \cite{Mellerud2021BitcoinTexas,McCook2022DriversEmissions,Carter2021BitcoinZero,Snytnikov2022FlareCapture,Jacobs2020InnovatorsPower,Vazquez2022FlaredOperations,OSTP2022ClimateStates} to take advantage of this nearly free energy source\footnote{This energy source can sometimes be negatively priced due to savings on flaring penalties \cite{Snytnikov2022FlareCapture} and production stoppages resulting from flaring caps \cite{Decker2021BitcoinField}.}. This process transforms methane emissions into carbon dioxide emissions through more efficient combustion\footnote{"Bitcoin mining (...) can burn the methane with a 99 percent efficiency, reducing substantially the risk of leakage into the atmosphere." \cite[41]{Oysti2021BitcoinConsumption}.}.

Assuming that methane is more harmful than \ac{CO2} in combating global warming\footnote{The \ac{IPCC} states that the \ac{GWP} "of methane are about 60, 28, and 5, respectively, for time horizons of 20, 100, and 500 years" \cite[15]{Brook2014WhyMix} (see also \cite{Oysti2021BitcoinConsumption,McCook2022DriversEmissions,Carter2021BitcoinZero,Vazquez2022FlaredOperations,Read2022GreenwashingIndustry}). However, the standard \ac{GWP} accounting method (\ac{GWP}100) has been challenged because it fails to consider \ac{GHG} emission \textit{rates}, leading to alternative metrics such as \ac{GWP}* \cite{Allen2018AMitigation}, which has an ``exculpatory" effect on some methane emissions, relative to \ac{GWP}100 \cite{Cusworth2022WhenMetrics}. These disagreements cannot be solved through scientific inquiry, as the underlying controversy is a political contention about attribution of responsibility, which depends on value judgments \cite{Cusworth2022WhenMetrics}. Statements about methane relative to other \ac{GHG}s should hence be made with care.} Bitcoin mining not only provides an additional stream of marginal income for energy companies per barrel of oil produced,\footnote{We should note however, that if an additional source of income prevents a drilling site from shutting down or stimulates additional hydrocarbon \textit{exploration}, the effect from Bitcoin-based gas flaring might not be net-decarbonizing (see also \cite{Roeck2022LifePlant,Rennie2021ClimateBitcoin,Jacobs2020InnovatorsPower} and \ref{subsec:second}).} but also adds to the load without compromising the existing energy supply. It furthermore provides a public service, by reducing the carbon footprint even where the total amount of gas extracted remains unchanged.\footnote{An additional incentive for oil and gas companies to permit bitcoin mining with flared gas is simply to get carbon emissions ``off their books," which in itself does not have a direct net decarbonizing effect. However, indirectly it does produce a decarbonizing impact due to the efficiency gains in combustion produced by transferring gas emissions to miners \cite{Rennie2021ClimateBitcoin}.} Indeed, methane-based bitcoin mining is reported to reduce \ac{GHG} emissions by 50\% to 63\% compared to traditional flaring, on top of making use of otherwise wasted energy \cite{Quirk2021CryptocurrencyInnovation.,Vazquez2022FlaredOperations,McCook2022DriversEmissions,Snytnikov2022FlareCapture}.\footnote{Similarly, Coinshares \cite{Coinshares2022TheImpact} argues that carbon emissions from flare mining are already negative and equivalent to 5\% of Bitcoin's global positive emissions already.}

This sort of mining, which is typically done off-grid (also not requiring additional transmission infrastructure), is usually \textit{not} flexible, and runs 24/7 (high uptime), as flared gas does not follow the intermittency patterns dictated by the sun or wind. On the other hand, gas operations may be shorter-lived.

The potential of methane-based mining is even recognized by the US \ac{OSTP}, which maintains that it is ``more likely to help rather than hinder U.S. climate objectives" \cite[24]{OSTP2022ClimateStates}. In this context, ``some suggest that the U.S. and Canada have enough flared natural gas to run the entire Bitcoin network" \cite[41]{Oysti2021BitcoinConsumption} (see also \cite{Snytnikov2022FlareCapture}).

\paragraph*{Others}

The above discussion covers some of the most notable applications of bitcoin mining in the \ac{RE} sector, but there are several other examples. \ac{BTM} hydroelectric mining is significant in the bitcoin mining market \cite{Rennie2021ClimateBitcoin}. Although its share is expected to decrease compared to wind and solar over time, it currently represents the primary \ac{RE} source for miners. Mining usually occurs with curtailed hydroelectric power in areas with excess hydro capacity (e.g., \cite{Liu2018HydropowerSuggestions}), and follows a higher uptime model than solar and wind energy (curtailment results from only a few hours of intense sun or winds \cite{Carter2021BitcoinZero}), leading to the usage of more modern \ac{ASIC}s \cite{BCEI2021BitcoinFuture}.

Other \ac{RE} sources are also employed, including biogas\footnote{Malfuzi et al. \cite{Malfuzi2020EconomicDemand} show that in countries with high electricity and natural gas prices, biogas mining is better suited for electricity generation and consumption in cryptocurrency mining, which they believe has environmental benefits.} and geothermal.\footnote{This is most notably experimented with in El Salvador. Kumar \cite{Kumar2022ReviewMining} finds that Bitcoin mining overcomes the energy transportation problem associated with geothermal energy generation, although \ac{ASIC} heating presents significant challenges in this context.}

Lastly, bitcoin mining may have additional applications at the intersection of energy and sustainability that extend beyond renewable penetration. The fierce competition in the bitcoin mining space\footnote{McCook \cite{McCook2022DriversEmissions} contends that bitcoin mining companies compete intensely in two dimensions: technological efficiency and managerial approaches. This is also demonstrated by the speed and minimal disruption with which large Chinese miners relocated their operations as the regulatory landscape became unfavorable.} makes mining a sector characterized by rapid innovation, a fact even acknowledged by bitcoin critics \cite{GhaebiPanah2022InvestmentMining}. Consequently, Imran \cite{GhaebiPanah2022InvestmentMining} posits that, by creating strong competition in computing power, bitcoin mining generates a positive externality: it encourages the development of chips that produce more computing power per unit of energy dissipated, outpacing Koomey's Law\footnote{The trend for the number of computations per joule of energy dissipated to double approximately every 1.5 years \cite{Koomey2011ImplicationsComputing}.}. Other positive externalities, such as arbitraging toward a global energy price, have also been identified \cite{Imran2018TheMining}.

\subsection{Business models}
\label{subsec:models}

Bitcoin mining offers potential growth opportunities at the intersection of the energy and cryptocurrency industries. This section explores various business models that utilize these synergies between the two sectors.

\begin{figure*}[h]
\centering
\begin{tikzpicture}[
  every node/.style={align=center},
  level 1/.style={sibling distance=45mm},
  level 2/.style={sibling distance=20mm},
  level 3/.style={sibling distance=5mm},
  level 4/.style={sibling distance=3mm},
  level distance=30mm, rotate=90]
\node {\textbf{Classification by}}
  child { node {Energy Source\\and Model}
    child { node {\ac{VRE}}
      child { node {Peak-shaving} }
      child { node {Valley-filling} }
      child { node {Load-building} }
    }
    child { node {Other \ac{RES}}
      child { node {Hydroelectric} }
      child { node {Nuclear} }
      child { node {Geothermal and others} }
    }
    child { node {Methane}
      child { node {Pay for the gas} }
      child { node {Pay for the equipment} }
      child { node {Others} }
    }
  }
  child { node {Location}
    child { node {On grid}
      child { node {\ac{FTM}} }
      child { node {\ac{BTM}} }
    }
    child { node {Off grid} }
  }
  child { node {Generator Relationship}
    child { node {Mere proximity} }
    child { node {Contracting} }
    child { node {Vertical integration} }
  }
  child { node {Pricing Model}
    child { node {Standard consumer} }
    child { node {\ac{PPA}}
      child { node {Standard} }
      child { node {\ac{DR} programs} }
    }
    child { node {Price responsiveness} }
  }
  child { node {Uptime}
    child { node {Low uptime} }
    child { node {High uptime} }
  };
\end{tikzpicture}
\caption{A typology of low-carbon mining models per defining factor.}
\label{diagram:typology}
\end{figure*}

\subsubsection{First or Last Resort Buyers}

\paragraph*{Buyer of First Resort}
Bitcoin mining can serve as not only a supplementary income source for energy producers but also as a primary demand source that pays more than alternatives, mainly selling to the grid \cite{Sarquella2022BitcoinAccelerator,Bastian-Pinto2021HedgingMining}. This may not necessarily displace other consumers because power plants may wait for years without selling energy to the grid due to connection queues, significantly harming profitability. Anticipating this issue, Bitcoin mining offers a consistent demand source, improving the economic viability of such projects and encouraging new plant installations \cite{Bastian-Pinto2021HedgingMining}. Furthermore, it can expedite construction and help adjust to unanticipated delays \cite{Bastian-Pinto2021HedgingMining}.

\paragraph*{Buyer of Last Resort}
Bitcoin mining can become profitable during periods of excess energy production or when transmission capacity is insufficient, resulting in extremely low energy prices \cite{BCEI2021BitcoinFuture,Rennie2021ClimateBitcoin,Sarquella2022BitcoinAccelerator}.

\subsubsection{Uptime}

\paragraph*{High Uptime Model} In a first-resort buyer scenario, the most efficient, latest-generation \ac{ASIC} miners are employed. These miners, which run almost 24/7, provide the lowest energy cost per hash \cite{Carter2021BitcoinZero}. Peak-shaving models are also high uptime.

\paragraph*{Low Uptime Model (Bottom-feeding)}
Less efficient Bitcoin miners, such as older \ac{ASIC} miners overtaken by newer machines in hash rate capacity, are typically pushed out of the market. The newer ASICs dilute the older ones' hash rate, pushing their break-even points to lower thresholds \cite{Oysti2021BitcoinConsumption}. However, when energy prices drop, previously unprofitable older \ac{ASIC}s become profitable to turn on intermittently. These miners can absorb excess energy from variable renewable energy generation facilities, functioning as ``\ac{ASIC} retirement homes" and offering an ancillary service to stabilize the grid and increase profitability \cite{Oysti2021BitcoinConsumption}.

\subsubsection{Miner Location}

\paragraph*{\ac{BTM}}

\ac{BTM} mining involves placing miners directly at the renewable energy \ac{RE} plant, reducing transmission and distribution costs \cite{Mellerud2021BitcoinTexas,BCEI2021BitcoinFuture,Roeck2022LifePlant,Hajipour2022AnMicrogrids,King2022BitcoinBehind-The-Meter}. This is contrasted to {\ac{FTM} mining.

\paragraph*{\ac{FTM}}

\ac{FTM} mining connects to the grid as a regular consumer, subject to the same prices and proportion of green to non-green energy as other customers \cite{Hajipour2022AnMicrogrids,Paez2021PolicyForce,King2022BitcoinBehind-The-Meter}.

\subsubsection{Pricing Model}

\paragraph*{Power Purchasing Agreement}

A \ac{PPA} offers a fixed price to the miner, providing stability and enabling forecasting. This can indirectly benefit the miner by facilitating external funding. A \ac{PPA} may be combined with an ``option" for the energy seller to require the buyer to turn off their \ac{ASIC}s, offering flexibility (see \ref{subsubsec:relationship}) in exchange for lower prices in non-peak demand periods as well as compensation for turning off the \ac{ASIC}s \cite{Braiins2021OptimizationsSources}. However, in areas with high renewable penetration or frequent extreme weather events, \ac{PPA}s may be priced above the median energy price \cite{Mellerud2021BitcoinTexas}.

A \ac{PPA} may take the form of:

\begin{description}
   \item \textbf{Standard \ac{PPA}:}  
Provides a fixed price and energy quantity for a specified time, with no unique considerations for a Bitcoin mining customer \cite{WB2021PowerEPAs}.

   \item \textbf{Ancillary service \ac{DR} programs:}   
   The miner locks in capacity at a certain price, selling an ``option on this capacity in the day-ahead ancillary service markets" \cite[42]{Mellerud2021BitcoinTexas}. \ac{CLR}s receive higher payments than \ac{NCLR}s. During emergencies, the grid operator ``deploys the load resources" by exercising the option to turn them off for up to a certain number of days per year \cite{Mellerud2021BitcoinTexas}.\footnote{Mellerud \cite{Mellerud2021BitcoinTexas} reports that the deployment of the load resources leads to an average uptime of 99.7\% in Texas, instead of 100\%. With this level of uptime, liquid cooling is required.}
\end{description}

\ac{PPA}-based mining is usually \ac{FTM} \cite{Mellerud2021BitcoinTexas}.

\paragraph*{Price-Responsiveness}

In a price-responsive model, the energy seller sells energy to the miner at the miner's break-even point when it is above the market price. If the market price is above the break-even point, the energy seller only sells to the market \cite{Mellerud2021BitcoinTexas}. This model may reduce energy prices paid relative to a \ac{PPA} because extreme price events do not affect the miner's price \cite{Mellerud2021BitcoinTexas}, and can also decrease other costs, such as cooling costs.\footnote{In Texas, the price-responsive model leads to an uptime of 85\%, where downtime coincides with the hottest periods of the summer. This minimizes the necessary expenses in liquid cooling \cite{Mellerud2021BitcoinTexas}.} From the energy seller's perspective, this model ``ensures" a local energy demand source at a minimum price, offering additional predictability.

Price-responsive mining is usually \ac{BTM} \cite{Mellerud2021BitcoinTexas}.

\subsubsection{Relationship between Miner and \ac{RE} Producer}
\label{subsubsec:relationship}
\paragraph*{Mere proximity}
The most basic relationship between a \ac{RE} producer and a miner involves no formal link, with the miner simply locating near the \ac{RE} plant \cite{Shan2019BitcoinCaiso,Roeck2022LifePlant}. This can be done to achieve ESG goals, take advantage of low energy prices during peak supply periods, or a combination of both. This simple relationship can result in efficiency gains \cite{Shan2019BitcoinCaiso}. However, it exposes the miner to energy price instability and does not give the producer control over the miner's operational status.

\paragraph*{Direct contracting}
 Increasingly, miners are establishing formal relationships with \ac{RE} producers \cite{Mellerud2021BitcoinTexas,Carter2021BitcoinZero}. A \ac{PPA} is usually established with the energy producer, providing pricing stability that the miner may prefer for its own reasons or to meet funders' (e.g. a bank's) concerns to release funding. In addition to the \ac{PPA}, the aforementioned option right is usually sold to the \ac{RE} producer \cite{Mellerud2021BitcoinTexas,Carter2021BitcoinZero}. This model offers the producer control over the miner's operational status, enhancing efficiency by allowing the energy supplier to occasionally influence demand and allocate energy more efficiently. However, energy consumption decisions may still deviate from optimal efficiency when exercising the option right is not possible or optimal, and transaction costs are not internalized.

\paragraph*{Vertical integration}
Though mostly theoretical to date, the possibility of a \ac{RE} producer directly engaging in mining \cite{Bastian-Pinto2021HedgingMining,Freier2023BitcoinGermany,Kraft2022TheSweden} or a miner entering the RE sector \cite{Kraft2022TheSweden,Soluna2019VerticalIntegration} (see also \cite{Handagama2022HowMiners}) has been considered. This business model requires knowledge diffusion and building expertise, as well as additional economic calculations on the energy producer's side. It allows for the continuous allocation of each specific electron to its most revenue-maximizing use (use it, store it, mine it \cite{BCEI2021BitcoinFuture}), potentially unlocking further efficiency gains.

Vertical integration may directly stabilize a \ac{RE} producer's income, acting as a \textit{real option} instead of an option contract \cite{Bastian-Pinto2021HedgingMining}. The miner could arbitrage between energy prices and bitcoin prices \cite{BCEI2021BitcoinFuture}. It may also ease capital availability and reputation issues for miners \cite{Rennie2021ClimateBitcoin,Roeck2022LifePlant,Kraft2022TheSweden,GDF2022Re:Assets}.

\subsubsection{Gas mining models}

Vazquez and Crumbley \cite{Vazquez2022FlaredOperations} identify two main business models for flare mining:

\paragraph*{``Pay for the gas"}

The miner pays for the collected gas used on the well site and keeps the mining proceeds \cite{Vazquez2022FlaredOperations}.

\paragraph*{``Pay for the equipment"}

In exchange for rental and service fees, the miner provides a fully equipped data center with generators to the hydrocarbon company, who keeps the mining proceeds \cite{Vazquez2022FlaredOperations}.

\paragraph*{Others}
Other models identified by Vazquez and Crumbley include ``mobile market hubs" to alleviate pipeline constraints \cite[5]{Vazquez2022FlaredOperations}.

\subsubsection{Portfolio greening}

Although beyond the scope of the current paper, a portion of the literature has concerned itself with the issues of proving that a bitcoin portfolio is ``green" and incentivizing green mining, regardless of whether this favors renewable penetration or not \cite{Ibanez2023DontBitcoin}. We briefly summarize existing proposals:

\paragraph*{\ac{RECs}, \ac{GOs}, and carbon offsets}
Miners can purchase \ac{RECs} and \ac{GOs} as a standard mechanism to reduce their carbon footprint \cite{Dogan2022AreMethod,Carter2021BitcoinZero}. While it is not possible to trace individual electrons through a grid with a varied energy mix, purchasing \ac{RECs} approximates showing that a given amount of energy has been sourced from renewables and indeed stimulates renewable buildout through a demand effect \cite{Carter2021BitcoinZero}. Both bitcoin miners and other buyers may also calculate the carbon footprint of their bitcoin holdings and purchase carbon offsets to compensate for it \cite{Cross2021GreeningOffsets,Coinshares2022TheImpact,Carter2021BitcoinZero,Read2022GreenwashingIndustry}.\footnote{Note however that Corbet et al \cite{Corbet2021Bitcoin-energyEvidence} find a negative correlation between bitcoin price and carbon credits. Similarly, they find no significant relationship between the bitcoin market and the largest green energy \ac{ETFs}. The authors interpret that this suggests that there are no positive externalities from Bitcoin to tackle climate change.} However, carbon offsets require knowledge of the broader energy mix of bitcoin mining with associated carbon accounting issues \cite{Carter2021BitcoinZero}, and there is significant controversy about their reliability \cite{Cross2021GreeningOffsets}. Furthermore, carbon accounting requires an investigation of the different forms of accounting for bitcoin mining emissions \cite{Ibanez2023CanSoK,WBD2022CanCross,Johnson2021GuidanceACcord} (see \ref{sec:consumption}).

\paragraph*{Incentive offsets}

Cross and Bailey \cite{Cross2021GreeningOffsets} argue that carbon offsets are neither reliable nor standardized, and that it is more profitable and effective to offset through incentives. Bitcoin mining is a zero-sum game, and any additional hashrate reduces rewards for existing miners. It follows that any investment in green hashrate acts as a Pigouvian tax on existing hashrate. Thus, a bitcoin investor may offset their carbon footprint entirely by co-investing in green bitcoin mining in proportion to their bitcoin holdings (in size \textit{and duration}) This also serves as an argument for authorities to provide tax breaks to green mining \cite{Shan2019BitcoinCaiso,SolunaComputing2022Re:Assets}.

\paragraph*{Others}
An alternative to \ac{RECs} and \ac{GOs}, which aim to indirectly certify that the miner used renewable energy in the mining process, and carbon offsets, which seek to compensate the investor's carbon footprint, is to purchase certificates for sustainably mined bitcoin. This offsets an investor's bitcoin holdings without requiring carbon accounting.\footnote{For instance, see Clean Incentive (https://www.cleanincentive.com/) and the Sustainable Bitcoin Protocol (https://www.sustainablebtc.org/).}

Other suggestions to encourage green mining include colored coins, or marked ``UTXOs from blocks discovered by mining pools with a known and sufficiently favorable energy mix" \cite[6]{Cross2021GreeningOffsets} (see also \cite{GhaebiPanah2022InvestmentMining}). This approach has significant disadvantages, such as breaking bitcoin fungibility and facing technical and accounting challenges \cite{Cross2021GreeningOffsets,Ibanez2023DontBitcoin}.

\section{Potential impact}
\label{sec:impact}

The impact of mining on \ac{RE} generation and energy grid management is already evident \cite{Cross2021GreeningOffsets, Mellerud2021BitcoinTexas, McCook2022DriversEmissions, Carter2021BitcoinZero, Snytnikov2022FlareCapture, Jacobs2020InnovatorsPower, Vazquez2022FlaredOperations}. However, the current scale of mining, whether based on renewable sources or not, is not large enough to significantly affect the global \ac{RE} sector \footnote{Shan and Sun argue that although bitcoin mining can mitigate the need for ramping capabilities in CAISO, it cannot manage the "neck" of the duck curve at sunset time at its current scale \cite[7]{Shan2019BitcoinCaiso}.}. Nevertheless, if \ac{PoW} mining were to be adopted on a much larger scale, the impact could be substantial \cite{WBD2022CanCross}. Some possible positive consequences of mining include:

\begin{figure*}
\centering

\includegraphics[width=0.6\textwidth]{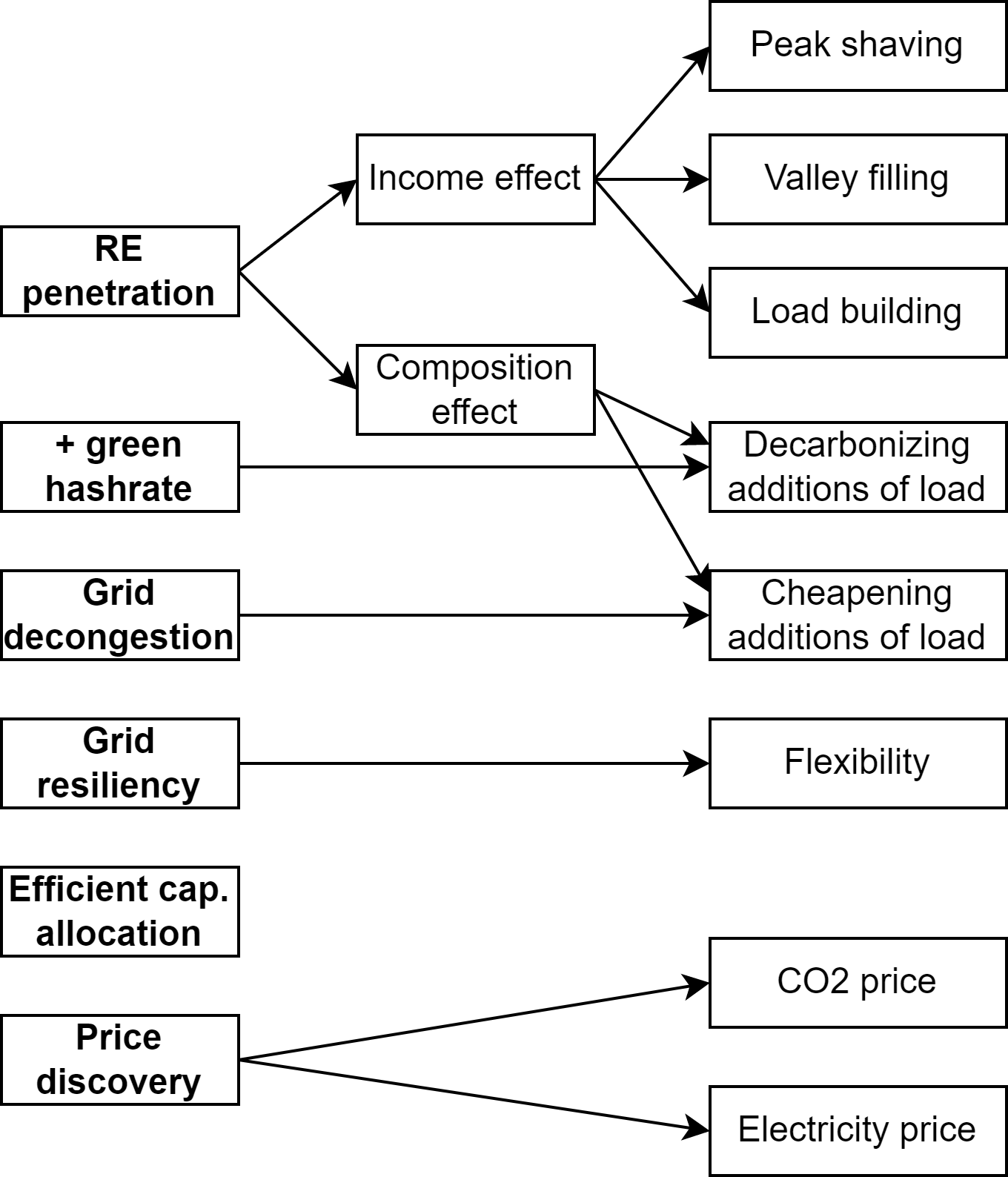}

\caption{Possible upsides of Bitcoin mining for decarbonization goals.}
  \label{diagram:upsides}
\end{figure*}

\paragraph*{Renewable penetration}

\ac{PoW} mining could potentially enable energy grid mixes with a higher contribution from renewable sources, leading to a decarbonizing effect. This could be attributed to several mechanisms:

\begin{description}
   \item \textbf{Income effect:} Bitcoin mining may subsidize or incentivize renewable capacity expansion on the margins by acting as a buyer of first and last resort \cite{Paez2022RFITransition, Shan2019BitcoinCaiso, BCEI2021BitcoinFuture}. Widespread adoption of Bitcoin mining as a complement to \ac{VRE}) generation could provide a substantial income source in addition to energy sales and serve as a stabilizing force. Mining can achieve three of the five primary demand-side management programs: peak-shaving, valley-filling, and load-building \cite{Attia2010MathematicalSolution}.

   \item \textbf{Composition effect:} Mining could enable the discovery of two key aspects of \ac{RE} markets: \textit{net-decarbonizing additions of load} and \textit{net cheapening additions of load}. The former refers to an increase in electricity demand that, counterintuitively, \textit{reduces} carbon emissions by making low-carbon energy sources more profitable and displacing high-carbon sources. The latter refers to increased energy demand that lowers prices rather than raises them, as upward pressure on prices due to increased demand is more than counterbalanced by increased renewable supply resulting from a surge in profitability.

   Both phenomena are considered plausible \cite{Carter2021BitcoinZero, Menati2022ModelingStudy}. For example, simulations by Lancium and IdeaSmiths, LLC suggest that introducing highly flexible data centers, such as Bitcoin mining facilities, to grids with an overabundance of wind power reduces \ac{CO2} emissions by decreasing reliance on natural gas for energy intermittency \cite{Rhodes2021ImpactsERCOT}. Instead of increasing generation (natural gas) during periods of stress, the market may decrease load, resulting in a net reduction of carbon emissions through near-zero carbon energy (ibid).

    Similarly, Nikzad and Mehregan estimate a 77.7\% reduction in the emission into the atmosphere of \ac{GHG} through the buildout of cogeneration projects of solar plants with cryptocurrency mining facilities \cite{Nikzad2022Techno-economicMining}. Dogan et al \cite{Dogan2022AreMethod} furthermore find that bitcoin clean energy and emission allowances are causally associated with bitcoin, in both volume and price, whereas miner revenues are negatively associated with carbon emissions (see also \cite{DiFebo2021FromSpillover}). Similarly, Menati et al \cite{Menati2022ModelingStudy} conclude that miner-driven additions of load could reduce energy prices.
    
    \item \textbf{Transmission decongestion:} Miners' portability (and scalability) means that they may be placed behind congested transmission nodes in the energy grid, further de-risking renewable buildout \cite{BCEI2021BitcoinFuture,King2022BitcoinBehind-The-Meter}.\footnote{Transmission/grid reinforcement, interconnection, and reserve capacities are expected to increase over time \cite{NorthSeaRegionProgramme2021FutureDevelopments,Joos2018Short-termGermany}, and hence node congestion might not constitute a permanent problem for grids \cite{Joos2018Short-termGermany}. Nevertheless, congestion management costs are increasing significantly at present \cite{Joos2018Short-termGermany}.}
    \end{description}

\paragraph*{Grid resiliency and reliability}

Widespread adoption of renewable mining could significantly enhance energy grid resiliency \cite{Menati2022ModelingStudy}. Grid resiliency refers to the grid's ability to adapt to rapid fluctuations and quickly recover from disruptions to supply, demand, or transmission capacity \cite{Shan2019BitcoinCaiso, Hajipour2022AnMicrogrids, Brook2014WhyMix}. It is of vital geopolitical importance \cite{GCGET2019ATransformation}. A more ``bitcoinized" grid would introduce an interruptible lever (\ac{CLR}) to regulate energy demand, allowing for improved reaction to various events, including rare ``black swan" events \cite{BCEI2021BitcoinFuture}\footnote{There are already instances of hashrate significantly dropping to ensure grid resiliency during winter storms \cite{McCook2022DriversEmissions}.}. This strengthens grid resiliency \cite{Paez2022RFITransition}, which is lower when renewable penetration is high \cite{Joos2018Short-termGermany}. Co-locating mining facilities with inflexible plants that produce stable loads effectively increases plant flexibility without altering generation practices.

Transmission decongestion also contributes to improved grid resiliency.

\paragraph*{Decarbonization through Green Hash Rate}

Mining can decarbonize not only by directly increasing \ac{RE} profitability and penetration, but also by indirectly penalizing high-carbon miners. Utilizing curtailed energy instead of grid-connected sources results in a clear net decarbonizing effect \cite{Shan2019BitcoinCaiso}. However, even when \textit{new} miners enter the market using \ac{RE}, the global hashrate is still decarbonized. As new miners increase the hashrate, the profitability of all other miners decreases, including those on the end-user side. This effect is due to the unique global zero-sum nature of the Bitcoin mining game. Therefore, whether existing mining infrastructure transitions to renewable sources or additional \ac{ASIC}s enter the market using renewable energy, the carbon intensity of mining is reduced.

\paragraph*{Mitigating Entrepreneurial and Government Errors}

While government promotion of \ac{RE} through subsidies and quotas can offer some benefits, it is also prone to exacerbating problems of entrepreneurial and government errors. Entrepreneurial errors may arise from market signal interference or subsidy delays, which can hinder ``creative destruction" \cite{Caballero1996OnDestruction} and the discovery of the most efficient processes. Government errors occur when they engage in ``picking winners and losers," allocating funds without a market process to determine expenditure efficiency, leading to capital misallocation. These issues were highlighted by Ludwig von Mises \cite{Mises1920EconomicCommonwealth} and Friedrich Hayek \cite{Hayek1945TheSociety} as the calculation problem and the information or knowledge problem.

By introducing a market-based mechanism to subsidize \ac{RE} \cite{BCEI2021BitcoinFuture}, Bitcoin mining as a complement to \ac{RE} generation can alleviate these issues while preserving price signals and the possibility of calculation. Mining can also help mitigate problems of capital allocation resulting from government action, for example, by capitalizing on overbuilt hydroelectric capacity. As Bitcoin acts as an ``apex predator" of energy \cite{Perrenod2022BitcoinPredator}, it may drive the discovery of the real value of electricity and potentially contribute to finding a global market price for carbon, a long-sought environmental objective \cite{Imran2018TheMining}.

\subsection{Difficulties in the market for \ac{PoW} mining}
\label{subsec:difficulties}

We identify a series of obstacles for \ac{PoW} mining-based decarbonization to succeed:

In general, miners face bitcoin price volatility in the short \cite{Vazquez2022FlaredOperations,Mellerud2021BitcoinTexas,Frumkin2020TheMining} and long \cite{Frumkin2021EconomicsEnergy} terms, as well as production volume volatility \cite{MellerUd2021BitcoinTexas}.\footnote{Miners traditionally do not hedge their investments but instead tend to "be long," hold their proceeds \cite{Frumkin2020TheMining}, and even borrow against their BTC holdings, exacerbating these circumstances.} However, price volatility can be hedged by shorting bitcoin futures, and new hedging instruments are emerging to hedge production volume volatility, such as difficulty derivatives, hash-rate derivatives, and hash-rate tokens \cite{Mellerud2021BitcoinTexas}. Moreover, there is the possibility to sell the hash rate \cite{Frumkin2020TheMining} (with green hash rate potentially sold at a premium). Additionally, profit margins tend to be low and have even turned negative for many miners due to low bitcoin prices and unexpected increases in hashrate \cite{Sarkar2022TopHighs,Freier2023BitcoinGermany}.

Other issues include the supply of \ac{ASIC}s. The mining market has recently experienced supply chain bottlenecks with a semiconductor shortage impacting downstream \cite{Perez2022TheShare,Redman2022GlobalTightens,Oysti2021BitcoinConsumption,Frumkin2021EconomicsEnergy}. Furthermore, the \ac{ASIC} market structure is highly concentrated, with Bitmain and MicroBT accounting for approximately 80\% of the market share \cite{Mellerud2021BitcoinTexas}. Although this can be considered a temporary stage in the market's evolution and potentially innovation-driving (see \ref{subsec:netnegative}), concentrated oligopolistic markets result in restricted supply (for both economic and technical reasons), creating a time lag between bitcoin price increases and the entry of new mining capacity. This limitation hinders the ability to execute large-scale demand-response \cite{Mellerud2021BitcoinTexas}.

\subsection{Internal Competition Between Renewable Mining Subsectors}
\label{subsec:competition}

We previously established that high-uptime waste (landfill, stranded, flare) gas mining and low-uptime wind/solar mining exploit near-free (if not actually free or negatively-priced) energy that would otherwise be challenging to use if not for bitcoin mining's unique characteristics. As a result, these activities may be most efficient when undertaken with ``outdated" \ac{ASIC}s pushed out of the market by hashrate increases due to new mining capacity entry. This implies interactions between the two markets: an increase in one's profitability may indicate greater demand for common inputs, subsequently decreasing the other's profitability. These interactions have not been thoroughly researched to date.

\subsection{Regulation and public outcry}
\label{subsec:regulation}

\ac{PoW} mining might not continue to exist at its current scale if its ``social license to operate" is revoked. \ac{PoW} bans, which have already been imposed \cite{Carter2021BitcoinZero} and considered \cite{OSTP2022ClimateStates,Paez2021PolicyForce} in many jurisdictions, pose a risk to the market's stability and sustainability. The unpopularity of mining in specific communities \cite{Mellerud2021BitcoinTexas} may trigger further unfriendly regulation.

The favorability of regulatory interventions is not just a matter of banning or allowing but also of the friendliness\footnote{This does not completely exclude forms of decarbonizing regulatory \textit{unfriendliness}. For instance, Roeck and Drennen \cite{Roeck2022LifePlant} argue that miners should be \textit{forced} to maintain a high renewable content to drive miners into greener areas (``carbon leakage").} of the regulation, its fine-tuning, and legal certainty\footnote{For instance, the abrupt removal of an electricity tax discount specifically for bitcoin miners in 2018 suddenly made mining almost entirely infeasible in Norway \cite{Mellerud2021BitcoinTexas}.} \cite{Carter2021BitcoinZero}. Miners tend to move to Bitcoin-welcoming jurisdictions\footnote{Mellerud \cite[60]{Mellerud2021BitcoinTexas} lists the following as friendly jurisdictions: Japan, South Korea, the Netherlands, Portugal, Switzerland, Georgia, Malta, and Singapore, as well as Wyoming, Florida, and Texas in the \ac{US}.

Specifically, with regards to Texas, Mellerud \cite{Mellerud2021BitcoinTexas} identifies a few key features of favorable market structure. First, deregulation, which entails few barriers, the existence of several electric providers competing for customers, and the ability for miners to negotiate directly with power plants without the need for an intermediary utility.}, but the continuity of a friendly environment remains always unknown. It is important to note that fossil energy subsidies, which artificially undercut the ability of \ac{RES} to sustain mining activity, should also be considered part of the regulatory environment \cite{Coinshares2022TheImpact}.

Naturally, regulatory friendliness may be partially endogenous, as a threatening political environment combined with sufficient sector growth may lead to the emergence of advocacy and lobbying efforts.

\subsection{Net negative impacts}
\label{subsec:netnegative}

Although it is possible that mining will lead to renewable buildout and hence \textit{net decarbonizing additions of load}, it is also possible that it will create a demand for additional conventional energy, which would increase \ac{GHG} \cite{Read2022GreenwashingIndustry,OSTP2022ClimateStates}. There have been isolated examples of this \cite{Oysti2021BitcoinConsumption}.\footnote{E.g., natural gas-based power plants that have been prevented from closing or re-ignited for bitcoin mining purposes \cite{Roeck2022LifePlant,Rennie2021ClimateBitcoin}. However, there is often more to these cases, e.g., when the mining facilities enable affording gas desulfurization equipment for the remediation of ash landfills from prior decades of coal mining activities \cite{Spence2022BillEP.288}. Furthermore, some of these miners do purchase RECs, which, in spite of their limitations, do benefit the \ac{RE} sector.} It is also conceivable for \ac{FTM} mining to merely use up existing renewable capacity, displacing \textit{other} consumers toward fossil energy,\footnote{Although \ac{BTM} mining does not display this problem, it has challenges of its own, as life cycle assessments are harder to conduct outside of the grid, which leads to obscurity \cite{Roeck2022LifePlant}.} despite miners' claims to be green due to the local grid mix (see theories of causality in \ref{paragraph:magnitudes}).

Although it is possible that mining will lead to renewable buildout and hence \textit{net decarbonizing additions of load}, it is also possible that it will create a demand for additional conventional energy, which would increase \ac{GHG} \cite{Read2022GreenwashingIndustry,OSTP2022ClimateStates}). There have been isolated examples of this \cite{Oysti2021BitcoinConsumption}.\footnote{E.g. natural gas-based power plants that have been prevented from closing or re-ignited for bitcoin mining purposes \cite{Roeck2022LifePlant,Rennie2021ClimateBitcoin}. However, there is often more to these cases, e.g. when the mining facilities are enabling affording  gas desulfurization equipment for the remediation of ash landfills from prior decades of coal mining activities \cite{Spence2022BillEP.288}. Furthermore, some of these miners do purchase RECs, which in spite of their limitations do benefit the \ac{RE} sector.} It is also conceivable for \ac{FTM} mining to merely use up existing renewable capacity, displacing \textit{other} consumers toward fossil energy,\footnote{Although \ac{BTM} mining does not display this problem, it has challenges of its own, as life cycle assessments are harder to conduct outside of the grid, which leads to obscurity \cite{Roeck2022LifePlant}.} despite miners' claims to be green due to the local grid mix (see theories of causality in \ref{paragraph:magnitudes}). 

Projects that utilize otherwise curtailed solar and wind energy may furthermore mine with grid electricity in periods when it is profitable to do so \cite{BCEI2021BitcoinFuture}, which on occasion may lead to positive \ac{GHG} emissions as well. We consider all these issues in more depth in sections \ref{subsec:second} and \ref{sec:discussion}.

\subsection{Scale}
In an electrified world, notwithstanding if Bitcoin's electricity consumption reached 1\% of the global electricity consumption, this might be insufficient for significant worldwide decarbonization even if the arguments in favor of Bitcoin's ``green" role are correct. On the other hand, this would mean that Bitcoin's electricity consumption would not manage to significantly aggravate climate change even if these arguments are wrong.

\subsection{Long-term equilibria}
\label{subsec:equilibrium}

While Bitcoin mining as a buyer of first resort may stimulate \ac{RE} buildout by providing a source of profits during interconnection queues, once electrification and decarbonization peak, those queues should progressively clear, meaning miners will have to consume from other sources.

\section{Alternative load resources}
\label{sec:alternatives}

Bitcoin mining is not the only flexible, interruptible, portable, and potentially nonrival source of energy load that may supplement grid decarbonization. In this section, we survey other activities that may play some of these roles.\footnote{``The short-term electricity overproduction can be partly mitigated by conversion to potential energy (hydropumped storage), chemical energy (batteries, hydrogen generation), or heat (aluminum smelters)" \cite[B]{Velicky2023RenewableBitcoin}.}

\subsection{Water desalination}
Water desalination is also a flexible, fully interruptible process that may be potentially supported with nonrival energy sources and act as a \ac{FLR} \cite{Atia2019Active-salinity-controlResource,BCEI2021BitcoinFuture}. Nevertheless, it is not as portable as Bitcoin mining as it requires pumping infrastructure, tanks and pumping scheduling systems \cite{Atia2019Active-salinity-controlResource}. Although some see desalination as a competitor to bitcoin mining \cite{Gasson2020Will2021}, there exist also proposals for bitcoin mining and water desalination as complementary infrastructures \cite{L392022HowPeople,Pearson2018ThisWell}.

\subsection{Water electrolysis for green hydrogen}
Green hydrogen\footnote{``Grey" hydrogen is produced from natural gas, "blue" hydrogen is also produced from natural gas but carbon is recaptured and/or reused, and ``green" hydrogen is renewable-energy based \cite{GhaebiPanah2022InvestmentMining,Snytnikov2022FlareCapture}.} production is also flexible, interruptible and potentially a good candidate for non-rival energy sources \cite{BCEI2021BitcoinFuture}.\footnote{Lund et al point out that not just hydrogen, but also synthetic methane, may be produced with excess energy \cite{Lund2015ReviewElectricity}.} Hence, the green dream of a ``hydrogen economy" \cite{Gupta2021Liebreich:Demand} may even spur a competitor to bitcoin mining itself \cite{GhaebiPanah2022InvestmentMining}.

However, electrolysis can be argued to be more expensive than mining, requiring additional equipment and infrastructure 
for storage and transportation \cite{GhaebiPanah2022InvestmentMining}. Electrolysis is also riskier and less lucrative \cite{GhaebiPanah2022InvestmentMining} than mining,\footnote{Ghaebi et al furthermore note that there is a frequent error of overoptimism about excess (free) \ac{RE} in flexible hydrogen electrolysis projects \cite{GhaebiPanah2022InvestmentMining}.} as well as less flexible.\footnote{Ghaebi et al \cite{GhaebiPanah2022InvestmentMining} note that electrolyzers are meant to run all year round to alleviate operation costs} Finally, the output of electrolysis is also less portable and storable than bitcoin \cite{Mellerud2021BitcoinTexas,Quirk2021CryptocurrencyInnovation.,McCook2022DriversEmissions,Carter2021BitcoinZero,Decker2021BitcoinField}.

Electrolyzers are furthermore unlikely to replace miners as marginal load providers, especially in a hydrogen economy. This is because such a system would require an energy consumption level many multiple above mining's projected peak, which could not be met with curtailed energy but in a fraction \cite{Liebreich2020Liebreich:Hydrogen,Liebreich2022GreenStrategy,Liebreich2020Liebreich:Side,Gupta2021Liebreich:Demand}. The sheer scale of a hydrogen economy would additionally entail transportation costs, storage costs and internal competition issues.\footnote{Between green hydrogen producers based on dedicated \ac{RES} and with curtailed \ac{VRE}, and between hydrogen and other \ac{RES}.}

Overall, hydrogen and Bitcoin present different profiles. First, electrolysis is almost entirely dependent on a single variable: hydrogen price. In contrast, mining is dependent not just on bitcoin's price, but also on other variables such as difficulty, hashrate, CAPEX, etc. \cite{GhaebiPanah2022InvestmentMining} Second, green hydrogen "specializes in seasonal demand flexibility", whereas Bitcoin mining is more suitable "for balancing unpredictable fast changes" \cite[9]{Mellerud2021BitcoinTexas}.

\subsection{\ac{CO2} removal}
Carbon dioxide removal is also a potentially flexible and interruptible activity that could be powered with non-rival energy sources, \cite{BCEI2021BitcoinFuture} although its profitability is still unclear. A key differentiator between \ac{CO2} removal and mining is that \ac{CO2} removal is a public good subject to the tragedy of the commons, which could make it an imperfect candidate for load balancing in the absence of very significant subsidies.

\subsection{Batteries}
\label{subsection:Batteries}
Batteries are a useful lever, both flexible and interruptible \cite{Mellerud2021BitcoinTexas}, and can solve part of the daily intermittency problem by balancing load \cite{Oysti2021BitcoinConsumption,Eid2021EnhancedLoad,BCEI2021BitcoinFuture,Brook2014WhyMix}. However, and even if their price is expected to continue falling, \cite{BCEI2021BitcoinFuture} batteries are expensive and lower \ac{ROI} in large scales \cite{Eid2021EnhancedLoad,Paez2022RFITransition,Brook2014WhyMix} because of physical limitations to the ability to store energy without dissipation \cite{Oysti2021BitcoinConsumption}. Moreover, they offer no additional benefit or profit other than flexibility itself \cite{Eid2021EnhancedLoad,Mellerud2021BitcoinTexas}, unlike Bitcoin, and they offer a smaller energy sink.\footnote{ESG concerns related to lithium set aside.} Nevertheless, there is evidence to suggest that batteries may provide an efficient complement to Bitcoin mining in the "right mix" \cite{Shan2019BitcoinCaiso,Eid2021EnhancedLoad,BCEI2021BitcoinFuture,Carter2021BitcoinZero,Winton2021SolarPower,Frumkin2021EconomicsEnergy,ARKInvest2021SolarBatteryBitcoin},\footnote{Eid et al \cite{Eid2021EnhancedLoad} find batteries to be inferior to bitcoin mining in terms of \ac{ROI}, but the combination of both to be a superior alternative to either activity in isolation, because of increases in profitability and optimizations to the batteries' "state of charge."

Frumkin \cite{Frumkin2021EconomicsEnergy} also notes that batteries, just like non-intermittent secondary energy sources, allow increasing uptime when it is not enough to break even on the mining CAPEX} especially in the upcoming electrified world, where excess energy will be additionally useful \cite{Oysti2021BitcoinConsumption}. Oeysti argues that "without bitcoin mining, renewables can provide only 40 percent of the grid’s demand," but a combination of "bitcoin mining, batteries and solar can provide 99 percent of the grid’s needs" \cite[40]{Oysti2021BitcoinConsumption} (see also \cite{BCEI2021BitcoinFuture}).\footnote{This is compatible with the IEA's projection that "to meet four-times the amount of hour-to-hour flexibility needs, batteries and \ac{DR} step up to become the primary sources of flexibility" \cite[177]{IEA2021NetSector}.}°

In this direction, consider that the energy consumption profiles of batteries are also different from miners': batteries only need energy for a few hours whereas miners may need it for longer periods. On the flip side, batteries can only supply energy for a limited time, whereas miners can technically be turned off indefinitely. This means that batteries are a better fit for some uses (e.g. on-site backup power) and bitcoin mining for others (e.g. sustained excess \ac{RE} over multiple days).

\subsection{Other flexible data centers and \ac{CLR}s}
Non-time-sensitive computation is not exclusive to \ac{PoW} mining. There is experimentation with the shifting of non-urgent computation tasks to match renewable generation peaks.\footnote{Such as uploading YouTube videos or adding new words to Google Translate \cite{Radovanovic2020OurBlows}. Note however that Google is a leader in its sector but this is not the norm in the data center industry \cite{Google2018MovingInsights}.}

Simulations have found flexible data centers to have indeed a net decarbonizing effect, as well as the effect to increase grid resiliency \cite{Velicky2023RenewableBitcoin}. Nevertheless, in both regards these centers are estimated to be inferior to cryptocurrency mining facilities \cite{Rhodes2021ImpactsERCOT} because of their comparatively lower flexibility \cite{Menati2022ModelingStudy} and efficiency \cite{Imran2018TheMining}. Similar is the situation of other \ac{CLR}s such as aluminum smelters \cite{Velicky2023RenewableBitcoin}

\subsection{Others}
Other alternatives include other forms of \ac{DR} programs, such as "sector coupling (power-to-gas, power-to-heat, and electric vehicles smart charging) (...) smart appliances in both commercial and residential buildings (...) industrial demand response (...) [and] load shedding schemes" \cite[10]{Mellerud2021BitcoinTexas} (see also \cite{GhaebiPanah2022InvestmentMining}). However, power-to-X solutions require "a meaningful probability of occurrence for low price hours to make it economically viable" \cite[5734]{GhaebiPanah2022InvestmentMining}, which limits their usefulness.

\section{Discussion}
\label{sec:discussion}

Throughout this article, we outlined the landscape of factors pertaining to Bitcoin's potential as a decarbonization tool. This allowed us to identify a few key areas for discussion.

\subsection{The problem of intermittency and profitability}
Although bitcoin can be mined with \ac{VRE}, the intense competition in the mining market raises the question of whether \ac{VRE}-based operations will be profitable at all. The concern is that intermittent mining could be continuously outperformed by mining based on non-intermittent energy sources. Intermittency leads to lower uptime,\footnote{Additionally, intermittency entails thermal cycling, which accelerated the degradation of silicon chip quality. However, there is little data on the extent to which this is a factor \cite{Braiins2021OptimizationsSources}.} which entails a longer time to recoup CAPEX, even under occasional negative energy prices conditions.\footnote{Currently, even extremely cheap electricity prices are not enough to make most operations viable with less than 60\% uptime \cite{Frumkin2021EconomicsEnergy}. 

} If, in turn, negative energy prices are very frequent, selling electricity to miners for free will not produce a significant benefit to a \ac{VRE} project's profitability and could hardly be considered a subsidy to renewable buildout. A risk-averse individual may consider this to be further aggravated by bitcoin's long and short-term price volatility.

Indeed the problem of uptime versus intermittency is common to all potential \ac{VRE} consumers, not just \ac{PoW} miners. In the worst-case scenario, this issue would mean that mining is neither genuinely interruptible nor a non-rivalrous consumer. However, a few observations should be made:

\begin{description}
   \item \textbf{Secondary power sources}: Intermittently-powered miners often rely on a secondary, more expensive electricity supply to underclock for in the absence of \ac{VRE} \cite{Braiins2021OptimizationsSources}. The combination of both may lead to a competitive \ac{LCOE}.\footnote{The \ac{LCOE} is the result of dividing the lifetime cost of building and operating a power generation asset by the kWh of energy that it produces \cite{Frumkin2021EconomicsEnergy}. \ac{LCOE} is not an all-encompassing metric, however, as it disregards system integration costs (costs of managing variability and uncertainty of energy output, e.g. by operating reserve and backup plants), which increase with \ac{VRE} penetration, hampering its political feasibility \cite{Joos2018Short-termGermany}.}
   
   \item \textbf{Low CAPEX miners:} Cheaper, older \ac{ASIC}s are less reliable and efficient, hence trading at a discount. As a result, their profitability depends comparatively more on OPEX than on uptime. Although indeed even a low CAPEX \ac{ASIC} is more profitable mining 24/7 than intermittently at the same electricity price, electricity prices are not constant, and this different profitability profile means that less efficient \ac{ASIC}s are more sensitive to electricity prices, potentially being more profitable under intermittent patterns.
   
    \item \textbf{Vertical integration:} Although selling negatively priced or near-zero electricity to a miner indeed does not add much to a generator's profitability, this does become the case when the miner and the generator are the same party. At low CAPEX, this possibility may come at the cost of little additional investment.
\end{description}

Additionally, if indeed 24/7 generation offers a substantial advantage over intermittent generation for mining, this may still pose a decarbonizing effect. Flare mining is a form of non-intermittent mining with a substantial decarbonizing effect, 
though it is unclear that it would consistently outperform \ac{VRE} mining. As it faces higher CAPEX, such as generators, together with scalability obstacles, flare mining may progressively slide toward higher-efficiency miners, not competing with \ac{VRE} mining. It is conceivable that coal plants may stay at minimum generation levels during negative price periods, but this entails additional OPEX (fuel) in comparison to \ac{VRE} mining. Finally, hydro mining (which eases an important capital misallocation issue in an environmentally friendly manner) faces important limits to growth in the long run, as hydroelectricity is not expected to grow as much as \ac{VRE} \cite{IEA2023Renewables2027}.

Ultimately, it is not impossible (though it is indeed unlikely, see \ref{subsec:second}) that the profit asymmetry between flexible mining with low-efficiency (and low-cost) rigs and uninterrupted mining with high-efficiency (and high-cost) rigs is so large that the former model is completely unprofitable. However, if the efficiency of newer rigs were indeed so high, this would depress the energy consumption of the Bitcoin network, mitigating the environmental concerns as well.

\subsection{Fairness of criticism and framing}

The plausibility of the opportunities discussed in this paper does not manage to \textit{guarantee} that \textit{every} addition of load will be carbon-neutral, or of the extremely flexible kind. It is also challenging to provide hard assurances concerning decarbonization timeframes. Similarly, the expectation that Bitcoin's energy consumption will eventually stop climbing and stabilize \cite{Mellerud2021BitcoinTexas,Carter2021BitcoinZero} is plausible, but not guaranteed. On the other hand, these requirements may be argued to be excessive, as no other industry is subjected to such extreme impositions. Consistency would require also criticizing other data centers for contracting \ac{RE} (e.g. Google's \cite{MarquesLima2022GoogleCentres}) that could have been destined for other uses.

It should also be noted that requiring new additions of load to be carbon-neutral presumes a particular theory of causality based on incumbency (see \ref{paragraph:magnitudes}), which to attribute impact to the "marginal load" may equivocate proximate causes and ultimate causes, assume a negative value judgment about Bitcoin, implicitly presuppose that older bidders of electricity have a more legitimate claim to the green share of electricity than younger ones (which needs to be justified), or incur in double-counting (attributing emissions to the marginal load and the incumbent load simultaneously). Regardless, the criticism that mining leads to an additional marginal load should also be accompanied by the admission that mining leads to an additional marginal revenue for \ac{RE} producers by increasing energy demand, which is often not the case.

Overall, Bitcoin's "social license to operate" should be subjected to reasonable requirements. A location-agnostic buyer of last resort that can protect a downside case in financial models and purchases otherwise-curtailed energy need not be required to exclusively purchase \ac{VRE} energy insofar it does not consume under peak demand and price conditions (which would incentivise fossil capacity expansion).

\subsection{A trend towards a perfect competition environment?}
\label{subsec:second}

It is plausible to expect mining-based additional loads to stimulate renewable buildout. A key factor in this regard is that bitcoin mining not only displays a near-perfect competition environment, but also is very likely to slide even further toward perfect competition over time. Perfect competition entails marginal cost sensitivity, which as established favors complementing \ac{VRE} generation. We now explore these claims:

\paragraph{Mining is already close to perfect competition in the present day}
Mining produces an entirely homogeneous product, namely bitcoin, which is secured through strong property rights \cite{Carter2021BitcoinZero}, and can be traded at near-zero transaction costs through technologies such as the Lightning Network. Although there are still some imperfections in the dissemination of market information, it displays an increasingly large number of buyers and sellers. Furthermore, there are non-increasing returns to scale (in fact, scale may have decreasing returns if fear of a 51\% attack is triggered), there are few barriers to entry (apart from the time to get a grid connection) or exit and there is near-perfect factor mobility (as shown by the China ban experience and seasonal pre-ban migrations within China \cite{McCook2022DriversEmissions,Carter2021BitcoinZero,Mankala2022AnImpact}). Empirically, the Bitcoin network also seems to follow the Pareto Principle or 80/20 rule, in another indication that it approximates a perfect competition state \cite{McCook2022DriversEmissions} (see also \cite{Oysti2021BitcoinConsumption}). The main differentiator is generally the cost of energy.

\paragraph{Mining's tendency toward perfect competition is posed to exacerbate}
This is a result of a series of factors:

\begin{description}
   \item \textbf{Increases in difficulty}: They are a result of the increasing hash rate and drive down profitability, meaning that only the miners with lower electricity costs are able to survive the upward adjustments \cite{Frumkin2021EconomicsEnergy}. Difficulty is expected to continue rising rapidly for the next years, in a clear trend for the average cost of mining a BTC to equal BTC price \cite{Frumkin2021EconomicsEnergy}.
   
   \item \textbf{The effect of halving}: Every four years, block rewards are cut in half, which along with difficulty adjustments pushes miners to the lowest possible input costs, especially to survive BTC-USD bear markets \cite{Carter2021BitcoinZero,Cross2021GreeningOffsets}. A fixed BTC supply, together with halving, additionally reduces the profitability of mining.
   
   \item \textbf{The expected clearing in the \ac{ASIC} supply chain}: In recent years, the bitcoin bull market coincided with a bottleneck in the production of \ac{ASIC}s. This is however clearing and expected to continue clearing over time \cite{Perez2022TheShare} (see also \cite{Redman2022GlobalTightens}).
   
   \item \textbf{Commodification of mining equipment}: Although we established that the \ac{ASIC} production market is currently highly concentrated, there is an extended expectation that mining equipment will commodity, especially with the decreasing rate of efficiency increases (see below). This is expected to drive CAPEX further down, with prices following production costs rather than bitcoin prices \cite{McCook2022DriversEmissions,Oysti2021BitcoinConsumption}.
   
   \item \textbf{Decreasing rate of increases in rig efficiency}: Hashrate increases reduce the profitability (and increase the cost-sensitivity) of mining, but rig efficiency increases counterbalance this. A necessary condition for Bitcoin to become a true flexible demand sink is a slowdown in efficiency increases.
   
   Since the beginning of the \ac{ASIC} mining era (2013), rig efficiency has been periodically improving by orders of magnitude. However, this is decelerating significantly \cite{Braiins2021OptimizationsSources}. Efficiency increases due to hardware miniaturization fall as quantum limits (and hence, increasing error rates) are approached. A lower rate of hash rate increases entails higher \ac{ASIC} lifetime, which raises the supply and the price-sensitivity of the hashrate. This reduces the marginal revenue of mining, which simultaneously limits total hashrate growth and exacerbates cost-sensitivity, entailing in turn a higher tolerance of downtime (intermittency) and hence additional suitability of mining for load balancing.

   Note also that, as the hardware playing field levels, competitive advantages are decreasingly obtained through purchases of newer rigs, and increasingly through more sophisticated \ac{PPA}s and by mining-\ac{VRE} co-location \cite{Braiins2021OptimizationsSources}.

   \item \textbf{Increasing access to credit}:  As the mining industry matures, there is a plausible argument that miners will get increasing (credit ratings and) access to credit to underwrite new renewable buildout. This is realistic as other data centers have already done this, including mining facilities.\footnote{See Aspen Creek Digital, https://acdigitalcorp.com/.}

    \item \textbf{Limits to the ability of bitcoin's price to increase}: Together with rig efficiency increases, the other factor increasing mining profitability and reducing cost-sensitivity is bitcoin price increases. The more BTC price grows, the lower the decarbonizing effect of the other trends identified in this article. However, BTC price cannot grow indefinitely in an exponential manner, meaning that as the digital asset achieves mainstream adoption and its price stabilizes, its suitability for load balancing will increase. Mellerud estimates that  "in the long-term, Bitcoin’s energy consumption will only continue to climb if the bitcoin price more than doubles every four years" \cite[61]{Mellerud2021BitcoinTexas}. Alternatively, miners would have to reduce the cost of IT by 50\% every 4 years, but even if CAPEX can come down (increasing the weight of OPEX), it cannot halve repeatedly.
   
\end{description}

The combination of these trends strongly suggests that miners' marginal cost will trend toward equalling their marginal income, and hence that the incentive to operate flexibly will significantly increase over time. In this context, it is reasonable to expect only or mostly near-free, free, or negatively-priced energy sources to lead to economic profits, and for miners to be encouraged to contract with newly-built wind and solar plants, meaning that a highly renewable scenario is more plausible with a large-scale mining scenario than otherwise. \cite{Quirk2021CryptocurrencyInnovation.}

There is a strong expectation for bitcoin mining is going to become ultra-competitive in a relatively short term \cite{Braiins2021OptimizationsSources}, with many miners going bankrupt in a bear market and selling their assets to other miners with lower energy prices -- all leading to a strong push for efficiency in a zero-sum game. 

\begin{figure*}
\centering

\includegraphics[width=\textwidth]{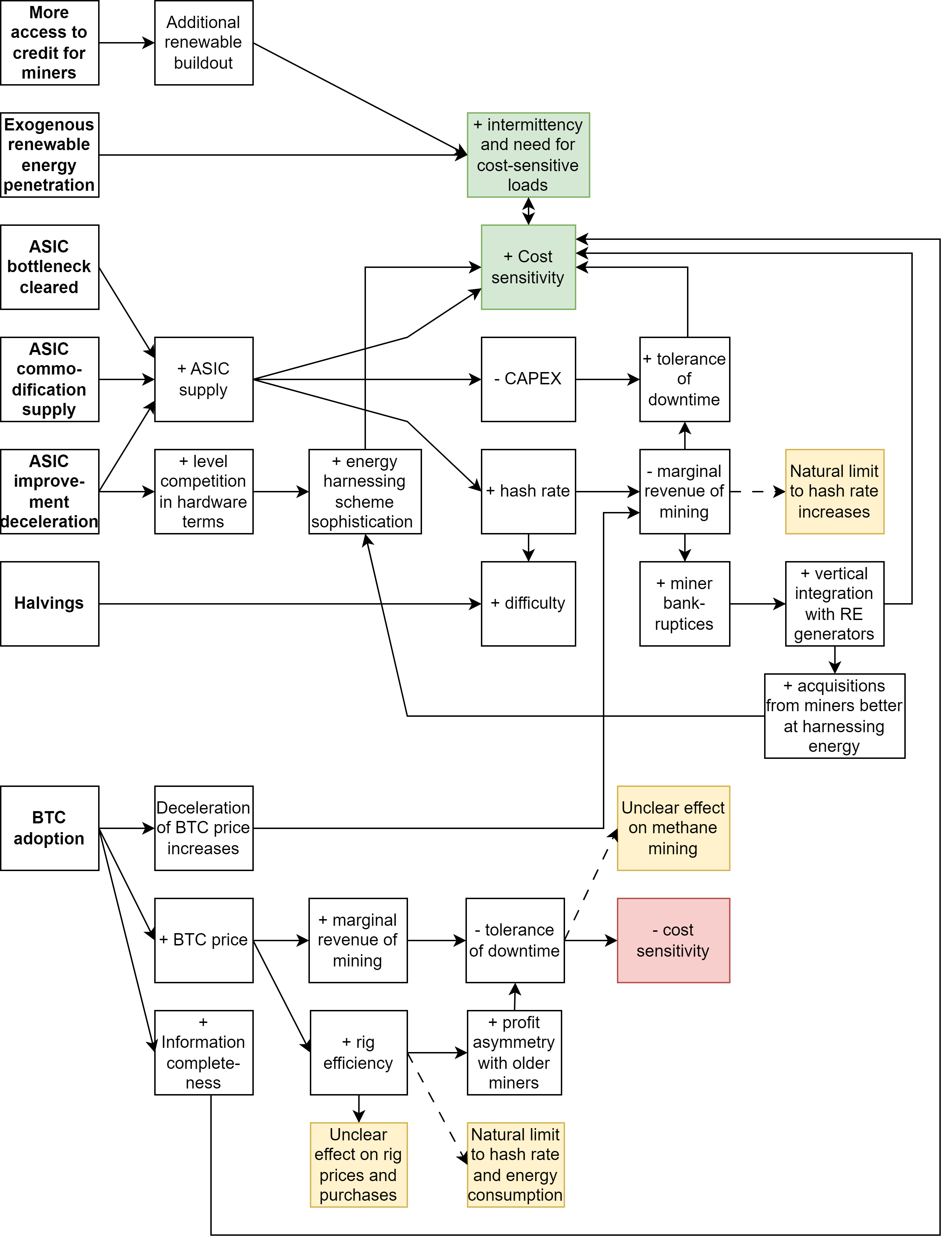}

\caption{Factors influencing Bitcoin mining's cost-sensitivity, with additional cost-sensitivity entailing more suitability for an ancillary services provider role.}
  \label{diagram:projections}
\end{figure*}

We observe that much of the disagreement regarding Bitcoin's potential for decarbonization seems to relate to how long the view that the proponent is willing to take is. A very immediate concern calling for immediate emergency degrowth in energy consumption requires a negative evaluation of Bitcoin's energy consumption. In turn, the expectation of Bitcoin to run only on \ac{BTM} intermittent \ac{RES} in a decade or two, but not entirely doing so in the interim, may lead to friendlier perspectives on this digital network.

Bitcoin's projected energy consumption may be seen as a huge proportion of \ac{RE} generation if only the next few years are considered, but not if an additional decade is allowed for.
In a similar direction, Imran \cite[12]{Imran2018TheMining} expects miners to act in an arbitrageur manner, first targeting "areas of energy surplus such as China, Canada, Norway [and] Iceland" and later moving into the renewable sector, looking for near-zero or even negative costs (see also \cite{McCook2022DriversEmissions}).

\subsection{Takeaways}

Bitcoin mining\footnote{Our results are focused on Bitcoin but partially apply to other \ac{PoW} blockchains as well \cite{Gundaboina2022MiningSource}, especially non-ASIC-resistant ones. Nevertheless, the reader should consider the difficulty of not only multiplying Bitcoin's current scale several times, but also competing \ac{PoW} \ac{DLT}s simultaneously matching this scale} displays a promising, though not yet entirely confirmed, potential to support \ac{RE} penetration.
This is relevant as it shows a tool that may be of aid toward grid decarbonization. 

Bitcoin mining is drawn to inexpensive sites for power sources. Although these can sometimes be carbon-intense sites, there is evidence to suggest that \ac{VRE} sources are preferred and that this trend will strengthen.\footnote{Additionally, general decarbonization efforts will positively impact Bitcoin's carbon-intensity profile insofar as it relies upon "the grid" \cite{McCook2022DriversEmissions,Carter2021BitcoinZero}}.
 Among renewable sources, while individual sites such as hydro or geothermal will continue to exist, solar and wind are gaining importance, given a cost and scalability advantage \cite{BCEI2021BitcoinFuture,Oysti2021BitcoinConsumption}.\footnote{The BCEI \cite[2]{BCEI2021BitcoinFuture} believes this to be especially true "for solar, a semiconductor technology, which has consistently declined in price by 20-40\%3 per doubling of cumulative capacity deployed."} 

Nevertheless, the argument that Bitcoin leads to decarbonization cannot be grounded just on Bitcoin being drawn to \ac{VRE} sources. For decarbonization, mining should also constitute a flexible load \textit{that does not add to peak demand}. With mining currently require a high uptime, this is not fully the case in the present day \cite{Braiins2021OptimizationsSources}.

Overall, and in spite of concerns about \ac{PoW} energy consumption, we have found a significant body of work suggesting that mining may be a powerful tool toward decarbonization. Further research in this direction (see \ref{subsec:future}) is thus required and encouraged. We moreover advise regulatory interventions to be formulated with extreme care, "as outright bans, punitive taxation or overly burdensome regulation" maybe have "the exact opposite of the desired effect by driving miners further into the jurisdictions where fossil fuels are heavily subsidized, thereby increasing emissions" \cite[18]{Coinshares2022TheImpact}.

Bitcoin mining for demand-response services is already a growing sector \cite{Mellerud2021BitcoinTexas}. This is not to say that there are no challenges, as several have been identified in this article. On a positive note, Vazquez and Crumbley \cite{Vazquez2022FlaredOperations} point out that all the advantages of bitcoin mining (e.g. interruptibility, flexibility, portability, etc.) are intrinsic \textit{technical} strengths. Instead, the disadvantages are mostly economic factors and contingent technical circumstances. An additional key advantage of Bitcoin miners is that they need minimal policy support (e.g. subsidies) to deploy themselves to supply renewable buildout.
 

\subsection{Future work}
\label{subsec:future}

With the entire sector being under-researched, many avenues for future research show promise. These include the exploration of specific business models for renewable-based mining, case studies showing profitability or the lack thereof empirically \cite{Freier2023BitcoinGermany}, scalability studies, ascertainment of the best geographic (and regulatory) locations for mining, the quantification of Bitcoin's positive and negative externalities, and more.

To guide public policy, we especially encourage rigorous end-to-end carbon accounting projects in the various renewable-based bitcoin mining niches, as well as, following Dogan \cite[12]{Dogan2022AreMethod}, the investigation of the contribution of \ac{PoW}-based cryptocurrencies "in determining the trade-off between renewable and non-renewable energy consumption," as well as the existence of potential non-linearities in this regard.

\subsection{Limitations}

This paper has not considered in depth other environmental impacts that go beyond GHG emissions, notably the impact of bitcoin mining in acidification, particulate emissions and smog formation, which have already been identified as areas for future research by others \cite{Roeck2022LifePlant}. Additionally, some areas of environmental impact have been considered more superficially than others (e.g. e-waste and noise pollution).

Furthermore, the reader should consider that this paper is not exhaustive of all the issues framing this discussion. For instance, an understanding of the explanation of bitcoin mining, Bitcoin's value and Bitcoin's value proposition are assumed and outside of the scope of this paper. Similarly, the 1.5 degree goal is taken as a standard reference framing the climate change debate and not necessarily advocated for \cite{Tol2013TargetsOverview}.

Finally, there are some obvious limitations that should also be taken into account. Most notably, research on the environmental impact of Bitcoin (even on Bitcoin itself) and especially on its impact on \ac{RE} is both still embryonic and fast-paced, meaning that significant findings may emerge after the release of this paper. This article also makes some elementary assumptions that are plausible but nonetheless contingent, such as that in the short run \ac{PoW}-based cryptocurrencies will not entirely cease to exist.

\subsection{Conclusion}

Bitcoin is indeed a network displaying a high electricity consumption. Nevertheless, this does not necessarily entail an equally high carbon footprint that is permanently sustained. There is evidence to suggest that bitcoin miners are unique energy buyers that may make bitcoin mining a \ac{FLR} for ancillary services provision. This has the potential for a net decarbonizing effect.

In this direction, it is not obvious that mining will necessarily be hyper-flexible. It is clear that Bitcoin-based decarbonization requires lower \ac{ASIC} utilization rates than the present day (and a higher relevance of electricity among total expenses), but advocating for Bitcoin-based decarbonization requires allowing for paths to net-zero that are not necessarily in a straight line.  

Although flexible loads may be instrumental in the fight against climate change, their usefulness naturally depends on the loads' willingness to be flexible. However, we observe a trend of the break-even value of miners going to zero, of renewable penetration (and, thus, curtailment) increasing, and even of bitcoin price volatility falling in the long run. This all suggests that Bitcoin may indeed be willing to be as flexible as decarbonization requires.

\section*{Acknowledgements}
\label{sec:acknowledgements}

We thank Troy Cross, Jens Strueker, Philipp Laemmel, Marcos Miranda, Paolo Tasca, Shaun Connell, Margot Paez, Florian Schemmerer, Jean-Philippe Vergne, Christian Ziegler and Elliot David 
for comments that greatly improved the manuscript.
J.I.I. and A.F. were supported by the University College London Centre for Blockchain Technologies. The authors are thankful to Energiequelle GmbH and Fraunhofer-FOKUS for their valuable comments and cooperation.


\section*{Conflict of Interest}

The authors declare that they have no known competing financial interests or personal relationships that could have appeared to influence the work reported in this paper.

\printacronyms

\printbibliography


\raggedbottom

\vspace{1em}



\end{document}